\documentclass[twocolumn,showpacs,preprintnumbers,amsmath,amssymb,floatfix,prb]{revtex4}
\usepackage{amsmath}
\usepackage{amssymb}

\usepackage{graphicx}
\usepackage{dcolumn}
\usepackage{bm}

\newcommand{\ie}  {{\it i.e.\ }}
\newcommand{\eg}  {{\it e.g.\ }}

\newcommand{\etal}{{\it et al.\ }}
\newcommand{\be}{\begin{equation}}
\newcommand{\ee}{\end{equation}}
\newcommand{\bea}{\begin{eqnarray}}
\newcommand{\eea}{\end{eqnarray}}

\begin{document}

\title{Two-electron lateral quantum-dot molecules in a magnetic field}

\author{M. Helle} \altaffiliation{Corresponding author: M. Helle
 before M. Marlo} \email{Meri.Helle@hut.fi} \author{A. Harju}
 \author{R. M. Nieminen} \affiliation{ Laboratory of Physics, Helsinki
 University of Technology, P. O. Box 1100 FIN-02015 HUT, Finland }

\date{\today}

\begin{abstract}

Laterally coupled quantum dot molecules are studied using exact
diagonalization techniques. We examine the two-electron singlet-triplet energy
difference as a function of magnetic field strength and investigate the
magnetization and vortex formation of two- and four-minima lateral quantum dot
molecules. Special attention is paid to the analysis of how the distorted
symmetry affects the properties of quantum-dot molecules.

\end{abstract}



\maketitle

\date{\today}

\maketitle

\section{INTRODUCTION}

The crossover from two-dimensional electron systems (2DES) to meso-
and nanoscale quantum dots (QDs) is an interesting subject. In the
infinite quantum Hall systems the actual arrangement of impurities or
disorder does not play a role, even if the presence of
disorder-induced localized states is vital for the Hall plateaus to
occur.~\cite{QHE} In the few-electron QDs, the type of disorder is
certainly an important issue. In the past, the majority of studies
have concentrated on highly symmetric parabolic QDs without disorder.
The rich spectrum of crossing energy levels as a function of magnetic
field and strong interaction effects are nowadays rather well
characterized for the cases with a symmetric confinement
potential.~\cite{kirja} Recently, the focus has turned to
understanding properties of QDs in less symmetric confinement
potentials.  For example, the rather simple far-infrared excitation
spectrum of a purely parabolic QD is nowadays well
understood,\cite{MaksymChakra} whereas the lowered symmetry introduces
new features in the spectrum whose interpretation is not
straightforward at
all.\cite{Hoch,MeriPRL,Veikko,Pfannkuche,Madhav,Magnusdottir,Ullrich,ChakraPietil_FIR_PRL05}
Moreover, the lowered symmetry also gives rise to modified ground
state properties such as level anticrossings and altered spin-phase
diagram as a function of magnetic field.~\cite{AriPRL,Brodsky}

After Loss and DiVincenzo proposition,~\cite{LossDiVincenzo} coupled
quantum dots have gained interest due to possible realization as
spin-qubit based quantum gates in quantum
computing.~\cite{Burkard_doubledot,SchliemannLossMacDonald,ScarolaDasSarma}
In addition to coherent single-spin operations, the two-spin
operations are sufficient for assembling any quantum computation.
Recent experiments have shown a remarkable success in characterizing
the few-electron
eigenlevels,~\cite{Lee,HuttelKotthausPRB05,ElzermanPRB03}
approximating relaxation and time-averaged coherence times and
mechanisms,~\cite{JohnsonMarcusNature05,KoppensKouwenhScience05} and
reading single-spin or two-spin
states~\cite{ElzermanNature04,HansonPRL05} of the QDs whereas the
coherent manipulation of spin systems remained out of reach until very
recent measurements on two-spin rotations.~\cite{PettaSciExp05}

In this paper we concentrate on two-electron quantum dot
molecules. These molecules consist of laterally, closely coupled
quantum dots. We treat correlation effects between the electrons
properly by directly diagonalizing the Hamiltonian matrix in the
many-body basis (exact diagonalization technique). This allows direct
access to the ground state energy levels and all excited states for
both spin-singlet and spin-triplet states. We study these levels as
well as singlet-triplet splitting and magnetizations as a function of
magnetic field and dot-dot separation. We also analyze the properties
of many-body wave functions in detail.

The magnetic field dependence of the ground state energy and
singlet-triplet splitting in non-parabolic QDs have attracted recent
interest.~\cite{EsaRECQD,Drouvelis,Drouvelis2,Drouvelis3,Szafran,Szafran2,Szafran3,Ugajin,AriPRL,Brodsky,Lee}
Magnetizations in QDs have been measured indirectly with transport
measurements~\cite{Oosterkamp} and recently with a direct technique
with improved sensitivity.~\cite{Schwartz} For both measurements,
semi-classical approaches cannot explain the results.  Also the
magnetizations of nanoscale QDs do not show non-equilibrium currents
and de Haas-van Alphen oscillations which are observed in 2DES and
mesoscopic QDs.~\cite{Schwartz2} In the nanoscale QDs the quantum
confinement and Coulomb interactions modify the system compared to the
2DES.~\cite{Schwartz} Theoretically, magnetization (at zero
temperature) is straightforward to calculate as the derivative of the
total energy with respect to magnetic field.  Magnetizations have been
calculated for a small number of electrons in a parabolic
QD,~\cite{MaksymMAG} in a square dot with a repulsive
impurity,~\cite{ShengXu} as well as for anisotropic
QDs,~\cite{Drouvelis} and for self-assembled QDs and quantum
rings.~\cite{Climente} The magnetization curves have been calculated
using density-functional theory for rectangular QDs~\cite{EsaRECQD}
and using the Hartree approximation for other types of non-circular
QDs.~\cite{Magnusdottir2} A tight-binding model for 10-100 electrons
in a single or two coupled QDs has been used to calculate
magnetization curves.~\cite{Aldea}

Calculations of vortices in QDs have also attracted much interest
lately.~\cite{HenriVortex,PeetersVortex,TorebladVortex,HenriVortex2,HenriVortex3}
Even if the vortices are not directly experimentally observable, they
reveal interesting properties of electron-electron correlations and of
the structure of the wave function. The nucleation of vortices in QD
systems could perhaps be observed by measuring magnetizations, where
each peak would correspond to one vortex added in the system. However,
the magnetization is difficult to measure for a small number of
electrons, especially with direct methods. Moreover, in a non-circular
symmetry, as in quantum dot molecules, and at high magnetic field
strengths the magnetization curves may become more complicated.

In our previous studies we have examined the properties of
two-electron, two-minima quantum-dot molecules (QDM) in a magnetic
field. The ground state as a function of magnetic field was found to
have a highly non-trivial spin-phase diagram and a composite-particle
structure of the wave function.~\cite{AriPRL} Also the calculated
far-infrared absorption spectra in two-minima QDMs revealed clear
deviations from the Kohn modes of a parabolic QD.  Surprisingly, the
interactions of the electrons smoothened the deviations instead of
enhancing them.~\cite{MeriPRL} In Ref. \onlinecite{MeriPhysica} we
briefly discuss some of the results of square-symmetric four-minima
QDM.

In this paper, we study in detail the properties of different QDMs.
Three different QDM confinements are studied thoroughly and their
properties are compared to parabolic-confinement single QDs. First, we
calculate measurable quantities such as energy eigenstates,
singlet-triplet splittings and magnetizations as a function of
magnetic field.  Secondly, non-measurable quantities, such as
conditional densities, vortices, total densities, and the most
probable positions are used to analyze the nature of the interacting
electrons in quantum states and also to analyze and understand the
properties of the measurable quantities.  This paper is an extension
to our previous calculations of QDMs.~\cite{AriPRL,MeriPhysica} We
study a two-minima QDM (double dot), a square-symmetric four-minima
QDM and a rectangular-symmetric four-minima QDM. The aim of this paper
is to study how the confinement potential affects the properties of
interacting electrons in a low-symmetry QD.

This paper is organized as follows: In Section~\ref{Model} we explain
how the quantum-dot molecules are modeled and what kind of basis we
use in the exact diagonalization method. We also discuss calculation
of magnetizations and the conditional single-particle wave function
which we use to locate the vortices and study the conditional
density. In the following four sections we present our results. In
Section~\ref{L0} we discuss properties of a single parabolic quantum
dot, and in Sec.~\ref{Lx5Ly0} we analyze the properties of a double
dot. In Sec.~\ref{Lx5Ly5} we present results for the square-symmetric
four-minima quantum-dot molecule, and finally in Sec.~\ref{Lx5Ly10}
the results for rectangular four-minima quantum-dot molecule.
The analysis of the results is presented in
Section~\ref{anaa}.  The summary is given in
Section~\ref{Summary}.

\section{Model and method}  
\label{Model}

\begin{figure}
\includegraphics*[width=0.55\columnwidth]{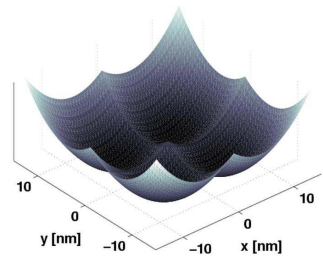}
\caption{Confinement potential of square-symmetric ($L_x=L_y=5$ nm)
four-minima quantum dot molecule.}
\label{pot}
\end{figure}

We model the two-electron QDM with the two-dimensional Hamiltonian
\begin{equation}
H = \sum _{i=1}^2\left ( \frac{ ( {- i {\hbar} \nabla_i}
-\frac ec \mathbf{A})^2 }{2 m^{*}} + V_\mathrm{c}({\bf
r}_{i}) \right ) +  \frac {e^{2}}{ \epsilon   r_{12} } \ ,
\label{ham}
\end{equation}
where $V_\mathrm{c}$ is the external confinement potential
taken to be
\begin{equation}
 V_\mathrm{c}({\bf r}) = \frac 12 m^* \omega_0^2 \min \left[
 \sum_j^M ({\bf r} - {\bf L}_j)^2 \right] \ ,
\label{Vc}
\end{equation}
where the coordinates are in two dimensions ${\bf r} = (x,y)$ and the ${\bf
L}_j$'s (${\bf L}_j = (\pm L_x, \pm L_y)$) give the positions of the
minima of the QDM potential, and $M$ is the number of minima. When
${\bf L}_1=(0,0)$ (and $M=1$) we have a single parabolic QD. With
$M=2$ and ${\bf L}_{1,2} = (\pm L_x,0)$ we get a double-dot
potential. We also study four-minima QDM ($M=4$) with minima at four
possibilities of $(\pm L_x,\pm L_y)$ (see Fig.~\ref{pot}).
 We study both square-symmetric ($L_x=L_y$) and rectangular-symmetric
($L_x \neq L_y$) four-minima QDMs. The confinement
potential can also be written using the absolute values of $x$ and $y$
coordinates as
\begin{eqnarray}
V_\mathrm{c}(x,y) &=& \frac 12 m^* \omega_0^2 \times \nonumber\\ &&
\left[ r^2 - 2 L_x |x| - 2 L_y |y| + L_x^2 + L_y^2 \right] \ .
\label{Vc_auki}
\end{eqnarray}
For non-zero $L_x$ and $L_y$, 
the perturbation to the parabolic potential comes from the linear
terms of $L_x$ or $L_y$ containing also the absolute value of the
associated coordinate.

We use the GaAs material parameters $m^*/m_e=0.067$ and
$\epsilon=12.4$, and the confinement strength $\hbar\omega_0=3.0$
meV. This confinement corresponds to the harmonic oscillator length of
$l_0 = \sqrt{\hbar/\omega_0 m^*} \approx 20$ nm. We concentrate on
closely coupled QDMs where $L_{x,y} \leq l_0$. The magnetic field (in
$z$ direction) is included in the symmetric gauge by the vector
potential $\mathbf{A}$.  The Hamiltonian of Eq. (\ref{ham}) is
spin-free, and the Zeeman energy can be included in the total energy
afterwards ($E_Z = g^*\mu_B B S_Z$ with $g^* = -0.44$ for GaAs). We
disregard the threefold splitting of each triplet state ($S_Z=0,\pm
1$) and consider only the lowest energy one ($S_Z=1$).

We drop the explicit spin-part of the wave function and expand the
many-body wave function in symmetric functions for the spin-singlet state
($S=0$) and anti-symmetric functions for the spin-triplet state ($S=1$).
\begin{eqnarray}
 \Psi_S({\bf r}_1,{\bf r}_2) = \sum_{i \leq j} \alpha_{i,j} \{ 
 \phi_i({\bf r}_1)\phi_j({\bf r}_2) \nonumber \\
 + (-1)^S \phi_i({\bf r}_2)\phi_j({\bf r}_1) \},
\end{eqnarray}  
where $\alpha_{i,j}$'s are complex coefficients. 
The one-body basis functions $\phi_{i}({\bf r})$ are 2D Gaussians. 
\begin{equation}
\phi_{n_x,n_y}({\bf r}) = x^{n_x}y^{n_y} e^{-r^2/2}, 
\end{equation}   
where $n_x$ and $n_y$ are positive integers.  The complex coefficient
vector $\alpha_l$ and the corresponding energy $E_l$ are found from
the generalized eigenvalue problem where the overlap and Hamiltonian
matrix elements are calculated analytically. The matrix is
diagonalized numerically.

\begin{figure}
\includegraphics*[width=0.6\columnwidth]{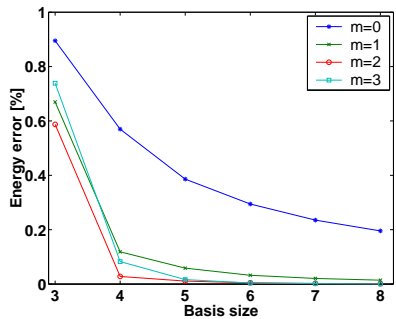}
\caption{Relative error in the energy of parabolic two-electron QD at
$B=0$ as a function of the basis size $n_x=n_y$ from 3 to 8, which
corresponds to around 40-2000 many-body basis functions in the
expansion. The relative angular momentum state of electrons is denoted
with $m$.}
\label{virhe}
\end{figure}

The basis is suitable for closely coupled QDs. At large distances and
at high magnetic field we expect less accurate results. The accuracy
may also depend on the symmetry of the state.  At \emph{zero} magnetic
field the \emph{parabolic} two-electron QD can be modeled with a very
good precision by expanding the basis (in a given symmetry) in
relative coordinates. In Fig.~\ref{virhe} we compare the energy of the
very accurate solution and the one using our basis (for a parabolic
QD) as a function of the basis size, where the maximum $n_x=n_y$
ranges from 3 to 8.
These values correspond to around $40-2000$
many-body configurations in the expansion. States with $m=0,1,2,3$
refer to different relative angular momentum states. The relative
error, even with the smallest basis studied $n_x=n_y=3$, is less than
one percent and decreases rapidly with the increasing basis size. The
greatest error is found for the $m=0$ state.

The magnetization can be calculated as the derivative of the total
energy with respect to magnetic field. It can be divided into to
parts, paramagnetic and diamagnetic,
\begin{eqnarray}
  M = - \frac{\partial E}{\partial B} = \langle \Psi|\frac {e}{2 m^* c} L_z +
g^* \mu_B S_z \ |\Psi \rangle \nonumber \\ - \frac{e^2}{8 m^* c^2} \langle \Psi|\sum_i r_i^2|\Psi \rangle B,
\label{extent}
\end{eqnarray}
where the former is constant as a function of magnetic field, for a
given angular momentum and spin state, and the latter depends linearly
on magnetic field. The diamagnetic contribution to the magnetization
is also a measure of the spatial extension of the ground
state.~\cite{ShengXu}

\begin{figure}
\includegraphics*[width=\columnwidth]{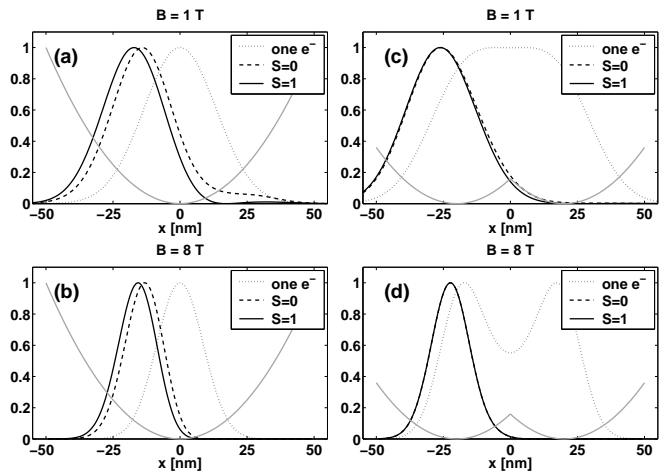}
\caption{One-body density (dotted line), two-body spin-singlet state
  (dashed line) and two-body spin-triplet state (solid line) along $x$
  axis. One of the electrons is fixed at the most probable position in
  the $x$ axis ($x^*$) and the conditional density is plotted for the
  other electron ($ |\psi_c({\bf r})|^2 =
  |\Psi_S[(x,y),(x^*,0)]|^2/|\Psi_S[(-x^*,0),(x^*,0)]|^2$).
  The peaks on the left-hand side also indicates the most probable
  position, therefore by reflecting the peak position to the
  right-hand side of the $x$ axis one can perceive the position of the
  fixed electron. (a) and (b) show the densities of parabolic QD at
  two different magnetic field values ($B=1$ and $8$ T), (c) and (d)
  represent two-minima QDM with $L_x=20$ nm. The confinement
  potential, $V_\mathrm{c}$, is plotted with gray color on each
  figure. }
\label{ddpot}
\end{figure}

Total electron density can be obtained by integrating one
variable out from the two-body wave function
\begin{equation}
n({\bf r}_1) = \int d{\bf r}_2 |\Psi_S({\bf r}_1,{\bf r}_2)|^2. 
\end{equation}
In practice we do not perform numerical integration. The density is
directly calculated in our diagonalization code where the required
matrix elements are calculated analytically.

We analyze the two-body wave function by constructing a conditional
single-particle wave function
\begin{equation}
\psi_c({\bf r}) = |\psi_c({\bf r})| e^{i \theta_c({\bf r})} =
\frac{\Psi_S[(x,y),(x_2^*,y_2^*)]} {\Psi_S[(x_1^*,y_1^*),(x_2^*,y_2^*)]},
\label{Eq_condwf}
\end{equation}
where one electron is fixed at position ($x_2^*,y_2^*$) and the
density ($|\psi_c({\bf r})|^2$) and phase ($\theta_c({\bf r})$) of the
other electron can be studied. One of the electrons is usually fixed
at the most probable position ($x_2^*,y_2^*$), but we also analyze
$\psi_c({\bf r})$ when the other electron is fixed at some other
position. The most probable positions of electrons (${\bf r}_1^*,{\bf
r}_2^*$) are found by maximizing the absolute value of the wave
function with respect to coordinates ${\bf r}_1$ and ${\bf r}_2$:
\begin{equation}
 \max_{{\bf r}_1,{\bf r}_2} |\Psi_S({\bf r}_1,{\bf r}_2)|^2
 \rightarrow {\bf r}_1^*,{\bf r}_2^* .
\label{rstar}
\end{equation} 
One should note that $|\psi_c({\bf r})|^2$ is \emph{not} normalized to
one when integrated over the two-dimensional space because it
describes the electron at position ($x,y$) on the condition that the
other electron is fixed at ($x_2^*,y_2^*$). Instead, $|\psi_c({\bf
r})|^2$ is normalized so that it equals to one when
$x=x_1^*,y=y_1^*$. Using the conditional single-particle wave function
we can study the conditional density $|\psi_c({\bf r})|^2$ and the
phase $\theta_c({\bf r})$.

To illustrate how the properties of the many-body wave function can be
examined with the conditional single-particle wave function, we
compare interacting two-body conditional densities to non-interacting
two-body densities in Fig.~\ref{ddpot}. The non-interacting two-body
density is the same as the single-particle density (up to a
normalization). We call it the one-body density hereafter. We plot the
one-body, two-electron singlet ($S=0$) and two-electron triplet
($S=1$) conditional single-particle densities along $x$-axis. The
other electron, in the two-electron systems, is fixed at the most
probable position ($x^*$) on the right-hand side of the
$x$-axis. 

\begin{figure*} 
\includegraphics*[width=2\columnwidth]{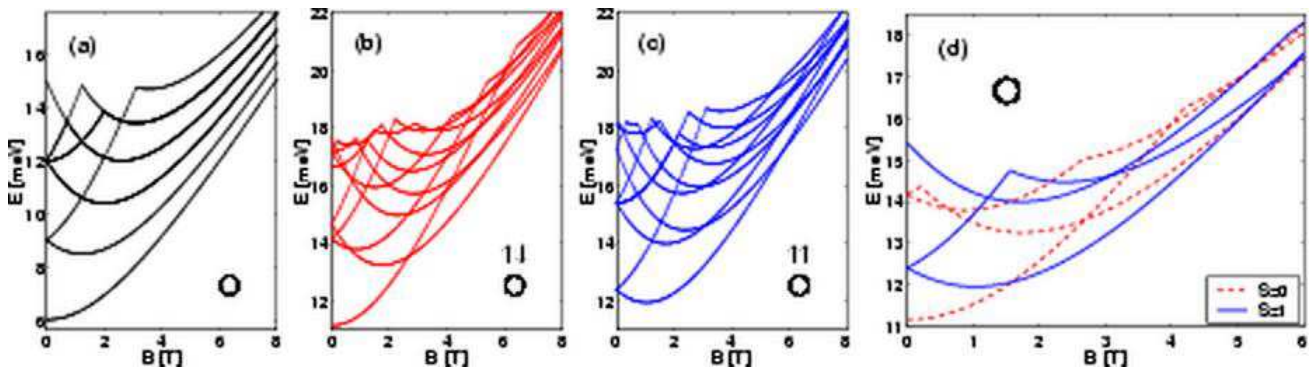}
\caption{Ten lowest energy levels of parabolic QD ($L_x=0, L_y=0$) as
a function of magnetic field of (a) non-interacting two-body, (b)
two-body singlet state ($S=0$) and (c) two-body triplet state
($S=1$). Some of the states in (a) are degenerate. (d) three lowest
energy levels of singlet (dashed line) and triplet (solid line) as a
function of magnetic field up to $B=6$ T for parabolic QD
($L_x=L_y=0$). In (b) the first singlet ground state
corresponds to angular momentum $m=0$ which changes to $m=2$ at $B
\approx 2.7$ T. In (c) the first triplet ground state is $m=1$ and it
changes to $m=3$ at $B \approx 5.8$ T. In (d) the first ground state
equals to $m=0$ singlet, then ground state is $m=1$ triplet followed
by $m=2$ singlet and $m=3$ triplet, where the latter is not visible as
a ground state in (d). Zeeman energy is included in the triplet
energies ($E_Z=-2 \times 12.7 B[T]$ $\mu $eV in GaAs).}
\label{elevLx0Ly0}
\end{figure*}

Fig.~\ref{ddpot} (a) and (b) show conditional densities of the single
parabolic QD at $B=1$ and $8$ T magnetic fields. The one-body density
is located at the center since no correlation effects push it towards
the edges of the dot. The peak of the triplet state is found further
at the edge of the dot than the singlet peak since Pauli exclusion
principle ensures that the electrons of the same spin are pushed
further apart than the electrons with the opposite spins. Notice that
the conditional density of the triplet state goes to zero where the
other electron is fixed, just before $x=20$ nm, whereas in the singlet
state there is a finite probability to find the electron around the
point of the fixed electron.

Fig.~\ref{ddpot} (c) and (d) show the same data for $L_x=20$ nm
two-minima QDM. In high magnetic fields and at large dot-dot
separations the difference between singlet and triplet densities
reduces. In QDMs, with a sufficiently large distance between the dots
and in high magnetic field also the one-body density localizes into
the individual dots (see Fig.~\ref{ddpot} (d)).

\section{Parabolic two-electron quantum dot ($L=0$)}
\label{L0}

We start our analysis from the single parabolic quantum dot. The
two-electron parabolic QD is studied extensively in the literature but
presenting results here serves as a good starting point for
understanding properties of quantum dot molecules.

\subsection{Energy levels}

Energy levels of the parabolic QD are plotted in Fig.~\ref{elevLx0Ly0}
as a function of magnetic field. Fig.~\ref{elevLx0Ly0} (a) shows
non-interacting two-body energy levels, (b) the ten lowest levels for
two-body spin-singlet states ($S=0$), and (c) for two-body
spin-triplet ($S=1$) states. In (d) three lowest singlet and triplet
levels are shown in the same plot. The non-interacting spectrum is
obtained by occupying two electrons in the Fock-Darwin energy
levels. The first eigenvalue at zero field equals two times ($N_e=2$)
the confinement potential ($\hbar \omega_0 = 3$ meV, $E_1(B=0)=3+3=6$
meV) and the second level represents one electron in the lowest
Fock-Darwin level and the other electron in the next one
($E_2(B=0)=3+6=9$ meV). Many non-interacting energy levels are
degenerate, also as a function of magnetic field. (In a less symmetric
confinement, the degeneracies are lifted). Due to degeneracies, only
six levels are seen in Fig.~\ref{elevLx0Ly0} (a). If the interactions
are included, the spectra become much more complicated and many more
level crossings are observed. One can also see how the energy scale is
modified. In the spin-singlet spectra the ground state energy is
almost doubled if the Coulomb interaction is included.

To see the crossing singlet and triplet states more clearly, we plot the three
lowest energy levels of spin singlet (dashed line) and spin triplet (solid
line) in Fig.~\ref{elevLx0Ly0} (d) up to $B=6$ T.  In a weak magnetic field the
ground state of the two-electron QD is spin-singlet ($S=0$), which changes to
spin-triplet ($S=1$) as the magnetic field increases and then again to singlet
and finally to triplet (at $B \approx 6.3$ T, not visible in
Fig.~\ref{elevLx0Ly0} (d)).

\subsection{Singlet-triplet splitting and magnetization}

\begin{figure}
\includegraphics*[width=\columnwidth]{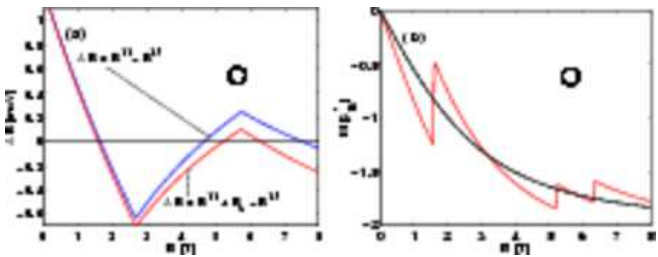} 
\caption{Triplet-singlet energy difference in (a) and magnetization in
(b) for parabolic QD ($L_x=L_y=0$). The lower curve in (a) shows the
singlet-triplet splitting with Zeeman energy included.  The smooth
curve in (b) represents the magnetization of two non-interacting
electrons and the other curve shows the magnetization for two
interacting electrons. At low magnetic field values, the ground
state is the angular momentum $m=0$ singlet, which shows as positive
values in the triplet-singlet energy difference of (a) and as a smooth
curve in (b). When the first transition from $m=0$ singlet to $m=1$
triplet occurs triplet-singlet energy difference changes its sign from
positive to negative and there appears a peak in
magnetization. Change of $m=1$ triplet to $m=2$ singlet and from $m=2$
singlet to $m=3$ triplet appear in the same way as peaks in
magnetization and changes of sign in triplet-singlet splitting.
Magnetization is given in the units of effective Bohr magnetons
$\mu_B^* = e \hbar/2 m^*$ ($\mu_{B}^* = 0.87$ meV/T for GaAs).}
\label{dELx0Ly0}
\end{figure}

In Fig.~\ref{dELx0Ly0} (a) we plot the energy difference of the lowest triplet
and singlet states up to $B=8$ T. Altering singlet and triplet states are also
seen in higher magnetic fields with a decreasing energy difference between
the states. However, if one includes the Zeeman term, the triplet state is
favored over the singlet state at high $B$. Therefore the system becomes spin
polarized.

The transitions between the states can also be examined from the
magnetization curves plotted in Fig.~\ref{dELx0Ly0} (b).  The
non-interacting magnetization is a smooth curve as no crossings are
present in the lowest energy level. The orbital angular momentum in
the non-interacting two-electron ground state does not change, and
thus only the diamagnetic effects are seen in the magnetization. The
non-interacting electrons have a smaller spatial extent of the wave
function compared to the interacting electrons.  Therefore, at low
fields, when the response is purely diamagnetic, the magnetization
curve of interacting electrons has a lower absolute value in
Fig.~\ref{dELx0Ly0} (b).  The magnetization curve of interacting
electrons shows abrupt increase of the otherwise smooth curve whenever
two levels cross (see also Fig.~\ref{elevLx0Ly0} (d)). The peaks in
magnetization are solely due to interactions.

\subsection{Wave function analysis \& vortices}

We will now analyze the two-body wave functions and study in more
detail singlet-triplet transitions in the single QD. The first
singlet-triplet transition can be be understood with the simple
occupation of the lowest single-particle states: In the singlet state
the two electrons occupy the lowest energy eigenstate with opposite
spins ($S=0$). As the magnetic field increases, the energy difference
between the lowest and the second lowest single-particle levels
decreases. (See non-interacting energy levels in Fig.~\ref{elevLx0Ly0}
(a)). At some point the exchange energy in the spin-triplet state
becomes larger than the energy difference between the adjacent energy
levels. Thus, the singlet-triplet transition occurs and the adjacent
eigenlevels are occupied with electrons of parallel spins ($S=1$).

However, the true solution of the two-electron QDM is much more
complicated than the occupation of single-particle levels and
inclusion of exchange energies. Interaction between the electrons
changes the situation drastically. This can already be seen by
comparing the single-particle energy levels of Fig.~\ref{elevLx0Ly0}
(a) to singlet and triplet energy levels in (b) and (c).  As a
signature of complex many-body features, many singlet-triplet
transitions are seen as a function of $B$. There are two trends
competing in the ground state of a quantum dot when the magnetic field
increases. The magnetic field squeezes the electron density towards
the center of the dot and the Coulomb repulsion of electrons increases
at the same time as the electron density is forced to a smaller
volume. At some point it is favorable to change to a higher angular
momentum ground state, which pushes electron density further apart and
reduces the Coulomb energy. Therefore as a function of
the magnetic field a series of different angular momentum states are
seen.

The altering singlet and triplet states can also be understood in
terms of composite particles of electrons and attached flux
quanta.~\cite{Jain} The starting point for understanding the ground
state changes and flux quanta is to consider the two-electron
parabolic QD (as discussed in this Section), which has an exact
solution for the wave function of the form
\begin{equation}
\Psi = (x_{12} + i y_{12})^m f(r_{12}) e^{-(r_1^2 + r_2^2)/2},
\label{Psi2QD}
\end{equation}
where $x_{12} = x_1 - x_2$, $y_{12} = y_1 - y_2$ and $r_{12} =
|\mathbf{r}_1 - \mathbf{r}_2|$ are the relative coordinates of the two
electrons, $m$ is the relative angular momentum and $f$ is a
correlation factor.\cite{AriPhysica,AriPRL} The zeros of the wave
function (vortices in relative coordinates) are placed on
$z_{12}=x_{12} + i y_{12}=0$ with a winding number given by the
relative angular momentum $m$.  In the first $S=0$ state the relative
angular momentum of electrons is zero ($m=0$). When the magnetic field
increases the ground state changes to spin triplet $S=1$, where the
relative angular momentum of electrons equals one ($m=1$) and in the
second singlet state $m=2$, and so on. With increasing magnetic field
the relative angular momentum of electrons increases and altering
singlet and triplet states are seen (if the Zeeman term is
excluded). The transitions occur because in the states with large $m$
the Coulomb repulsion becomes smaller at the cost of higher
confinement and kinetic energies. As the increasing magnetic field
squeezes electrons to a smaller area, it is favorable to move to
larger $m$ to minimize the total energy. One should note that with
even $m$ the spatial part of the total wave function is symmetric and
therefore the spin part should be antisymmetric ($S=0$). With odd $m$
the spin part is symmetric ($S=1$).

\begin{figure}
\includegraphics*[width=\columnwidth]{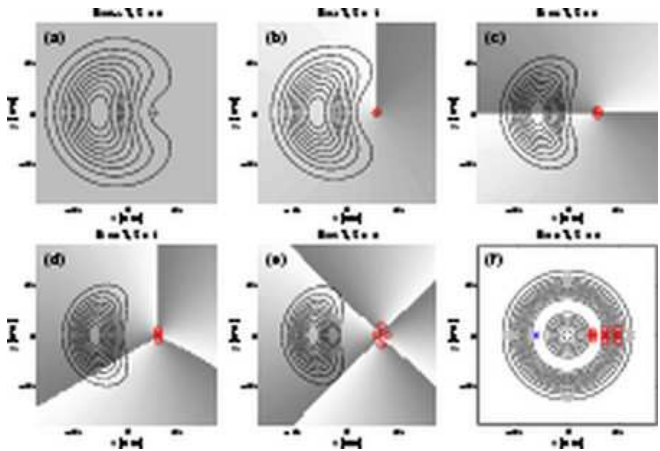} 
\caption{(a)-(e) contours of \emph{conditional} density
$|\psi_c(x,y)|^2$ and phase of the conditional wavefunction
$\theta_c(x,y)$ in gray-scale for parabolic QD ($L_x=L_y=0$). (White
equals $\theta_c=0$ and darkest gray $\theta_c=2\pi$). Magnetic field
value and the spin type are plotted on top of each figure. The plus
sign ($+$) indicates the position of the fixed electron and small
circles indicate the positions of the vortices. In (f) contours of
\emph{total} electron density of the three-vortex triplet state are
plotted in the background and positions of the vortices are solved
when the fixed electron is in three different positions. Fixed
electron is marked with the plus sign and vortices with circles.  The
most probable position is marked with a star, on the left-hand side
for clarity.}
\label{vortexLx0Ly0}
\end{figure}

When the angular momentum increases, the increased rotation induces
vortices in the system. As we have a many-body system, the rotation is
a correlated motion of electrons and can be studied in the relative
coordinates of electrons.  Vortices can be found by locating the zeros
of the wave function and studying the phase of the wave function when
going around each of the zeros. As the vortices are seen in the
relative coordinates, and are not visible in the density, we examine
the conditional single-particle wave function $\psi_c({\bf r})$ (of
Eq. (\ref{Eq_condwf})), where one electron is fixed in the most
probable position (on the $x$ axis, the system is circular
symmetric). The vortices are seen in the zeros of $\psi_c$.  When the
phase part, $\theta_c$, is integrated around a closed path encircling
the zero, we obtain the winding number of the vortex ($\oint
\theta_c({\bf r}) d{\bf r} = m 2 \pi $).

In a parabolic two-electron QD vortices are automatically attached on
top of the electrons, where the relative angular momentum $m$ equals
the winding number of a vortex. In a less symmetric potential the
center of mass and relative variables do not decouple and one would
expect more complicated structures as can be seen in later sections.
The simple form of the wave function in Eq. (\ref{Psi2QD}) is due to
separation of the center of mass and relative coordinates in the
parabolic confinement.

Calculated vortices and conditional densities of the single QD are
shown in Fig.~\ref{vortexLx0Ly0}. The contours show the conditional
electron density, $|\psi_c|^2$, and the gray-scale background marks
the phase of the conditional wave function, $\theta_c$, where the
white color equals $\theta_c=0$ and the darkest gray $\theta_c=2
\pi$. The positions of the vortices are marked with circles (o), and
the other electron is fixed at the most probable position (${\bf
r}_2^*$) shown with a plus sign ($+$). The lines of dark gray and
white borders correspond to a sudden phase change of $2 \pi$ if the
line is crossed. The number of flux quanta attached to the electron
(or the winding number of a vortex) can be determined by going around
the fixed electron position and calculating the total phase change (or
counting the lines crossed in the figure).

In Fig.~\ref{vortexLx0Ly0} (a) the phase is constant (no vortices and
no relative angular momentum) and the the probability density of the
other electron is located on the left side because of the Coulomb
repulsion. In (b) we find one vortex as one border of white and gray
is crossed when the fixed electron is encircled. In (c) we find two
vortices, in (d) three vortices and in (e) four vortices.
Fig.~\ref{vortexLx0Ly0} (a) corresponds to the first singlet with
relative angular momentum $m=0$, (b) shows the first triplet with
$m=1$, (c) the second singlet with $m=2$ and (d) the second triplet
with $m=3$. Fig.\ref{vortexLx0Ly0} (e) would be the third singlet
state but this is not a ground state if the Zeeman term is
included. The conditional density shows how the electron localizes to
a smaller area when the magnetic field increases. We also notice the
enhancement of interaction at high $B$ where the density contours are
contracted compared to the low-field conditional densities.

\begin{figure}
\includegraphics*[width=\columnwidth]{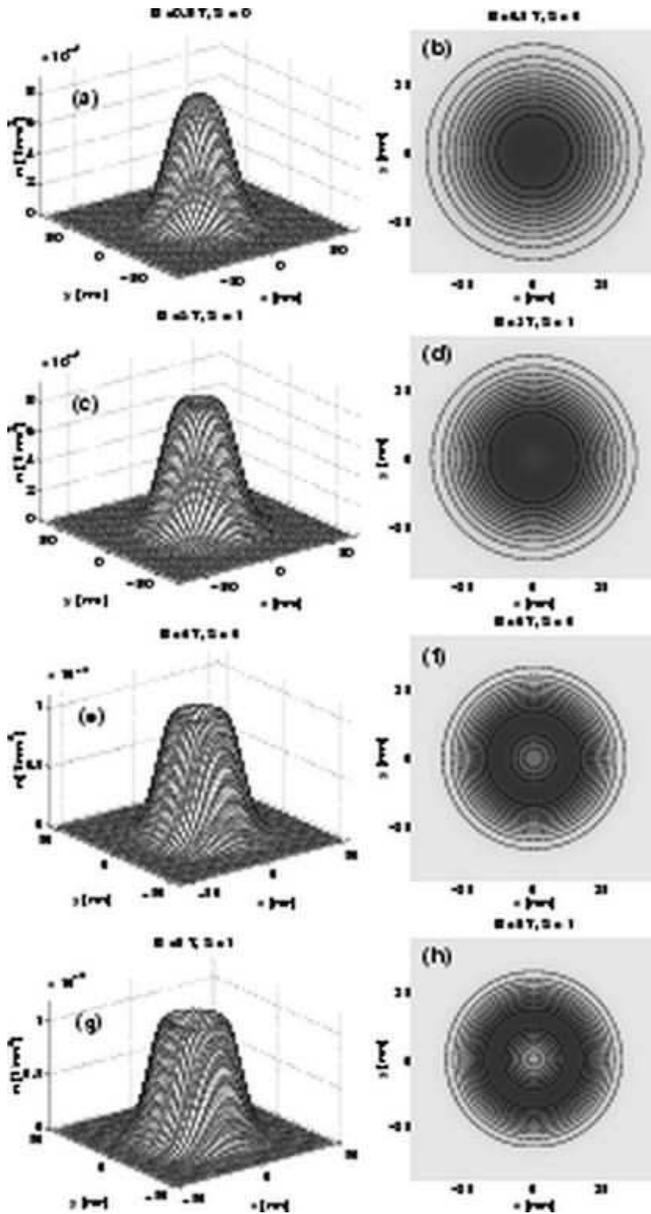} 
\caption{Total electron density of the ground state at different
magnetic field values for parabolic ($L_x=L_y=0$) QD. In the left
column are densities in 1/nm$^2$ and on the right column are contours
of the densities. Dark regions in the contours mark high density and
brighter regions low density. Notice that in the densities the scale
in the $x$ and $y$ axes is changing and the height of the peak is
increasing with magnetic field. The range of axes in the contour plots
is kept fixed. The magnetic field value and the spin type of the
ground state is plotted above each sub-figure.}
\label{densLx0Ly0}
\end{figure}

\begin{figure}
\includegraphics*[width=0.6\columnwidth]{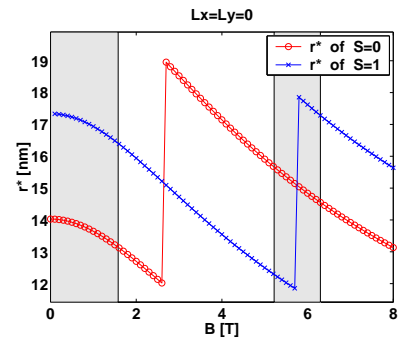}
\caption{Most probable position in nm of singlet ($S=0$) and triplet
($S=1$) states for parabolic ($L_x=L_y=0$ nm) QDM. The magnetic field
region where singlet is a ground state is marked with gray background
color.}
\label{rstarLx0Ly0}
\end{figure}

The conditional density and phase are much more sensitive to the basis
size than \eg energy eigenvalues. In a parabolic QD the vortices
should appear exactly on top of the fixed electron. However, in the
two-, three- and four-vortex plots, in Fig.~\ref{vortexLx0Ly0}, the
vortices are slightly displaced from the fixed electron position.  The
finite basis expansion does not result in exactly correct vortex
picture. On the other hand, the problem is not very serious because
the error in energy is not large and vortex positions are not
experimentally observable and we also get correct winding numbers
(angular momentum) if the pinned electron is encircled with a large
enough radius.

Vortex dynamics can be studied if we change the position of the fixed
electron. In Fig.~\ref{vortexLx0Ly0} (f) we show the positions of the
vortices in three different places of the fixed electron for the
three-vortex triplet state. The total electron density is shown in the
background (see also Fig.~\ref{densLx0Ly0} (g) and (h)).  In a
parabolic QD the vortices should be on top of the fixed electron no
matter where it is pinned. Our calculations result wrongly a small
offset of vortices from the fixed electron.  Even if the vortices are
not exactly on top of the fixed electron, and actually have a greater
offset at greater distances from the origin, they are seen to follow
the fixed electron if its position is changed.  This is a signature of
the composite particle nature of electrons and vortices.

\subsection{Total electron density and the most probable position}

In Fig.~\ref{densLx0Ly0} we plot ground state total electron densities
of the parabolic QD. The magnetic field values are the same as in
Fig.~\ref{vortexLx0Ly0} except that the four-vortex solution is not
plotted since it is not a ground state.  The range of both $x$ and $y$
axes and also the density in the $z$ axis changes from (a) to (g)
whereas in the contours the range of $x$ and $y$ is kept fixed. Dark
regions in the contours mark high density and bright regions low
density.  As a function of magnetic field there forms a minimum in the
center of the dot as the Coulomb repulsion forces electrons further
apart. Note that even if there is a minimum in the total electron
density, this is not a vortex. Vortices, in the composite particle
picture, follow moving electrons and are seen in the relative
coordinates of electrons. They describe the correlated motion of
electrons and are not visible in an averaged-out quantity such as
density.

Another way to study the nature of the changing ground states is to
plot the most probable position (see Eq. (\ref{rstar})) as a function
of magnetic field.  Fig. \ref{rstarLx0Ly0} shows the most probable
position (${\bf r^*}$) for both spin-singlet and spin-triplet states
as a function of magnetic field. Gray background color indicates the
region of magnetic field where the singlet is a ground state and white
background color corresponds to magnetic field regions where the spin
triplet is a ground state. Each jump in the curves corresponds to a
change in the angular momentum. When the singlet changes from $m=0$ to
$m=2$ state at $B \approx 2.7$ T the most probable position jumps to a
higher value as well. See also energy level crossings in
Fig.~\ref{elevLx0Ly0} (b). The outward relaxation, due to the increase
of angular momentum, can be also seen in the density.  Similar
dependence is seen for the triplet state. First the most probable
position decreases due to contracting electron density and then, at
some point, it is favorable to move to a higher angular momentum state
which relaxes the electron density outwards. In higher magnetic field
we would see a sequence of transitions between increasing angular
momentum states.

\section{Two-minima quantum dot molecule ($L_x \neq 0$)}
\label{Lx5Ly0}

\begin{figure}
\includegraphics*[width=\columnwidth]{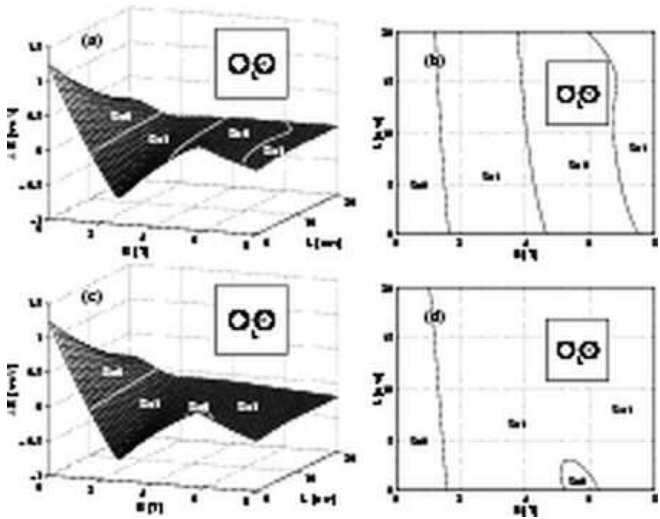}
\caption{Triplet-singlet energy difference ($\Delta E = E^{\uparrow
\uparrow} - E^{\uparrow \downarrow}$) as a function of magnetic field
in two-minima quantum dot molecule. The energy difference is plotted
as a function of dot-dot separation and magnetic field in (a) without
Zeeman energy and in (c) with the Zeeman energy included ($\Delta E =
E^{\uparrow \uparrow} + E_Z - E^{\uparrow \downarrow}$). In (b) and
(d) the ground state regions of the singlet and triplet states are
plotted as function of $B$ and $L$, without and with Zeeman energy,
respectively.}
\label{ELx5Ly0}
\end{figure}

\subsection{Singlet-triplet splitting as a function of $L$}

In this section we study two laterally coupled quantum dots.  In a
two-minima QDM, or double dot, we study the changes in the ground
state spectrum when two QDs, on top of each other, are pulled apart
laterally. In Fig.~\ref{ELx5Ly0} (a) the energy difference of the
lowest triplet and singlet states is plotted as a function of the
inter-dot spacing and magnetic field. At $L=0$ we have a single
parabolic QD and the curve coincides with Fig~\ref{dELx0Ly0} (a). When
$L \neq 0$ we have a double dot. Let us now examine some general
trends of the triplet-singlet energy difference as a function of
dot-dot separation. At small magnetic field the ground state is a spin
singlet, then triplet, and again singlet as in the single QD, but the
transition points change and the energy differences are smaller at
greater distances between the dots than in the single QD. The
transition points and regions of singlet and triplet states are
plotted in Fig~\ref{ELx5Ly0} (b). We can also note that all transition
points are shifted to lower $B$ at large distances between the dots.
If the Zeeman energy, that lowers the triplet energy, is included in
the total energy the second singlet state disappears at greater $L$ as
can be seen in Fig.~\ref{ELx5Ly0} (c) and (d). The second singlet is
only seen in a small region with very closely coupled QDs ($L \lesssim
2.5$ nm). Therefore, subsequent singlet states after the first one are
not seen in the double dot if $L \gtrsim 2.5$ nm.  Similar results are
seen in anisotropic QDs where the parabolic confinement of a single QD
is elongated continuously to a wire-like confinement.\cite{Drouvelis}

\begin{figure*}
\includegraphics*[width=2\columnwidth]{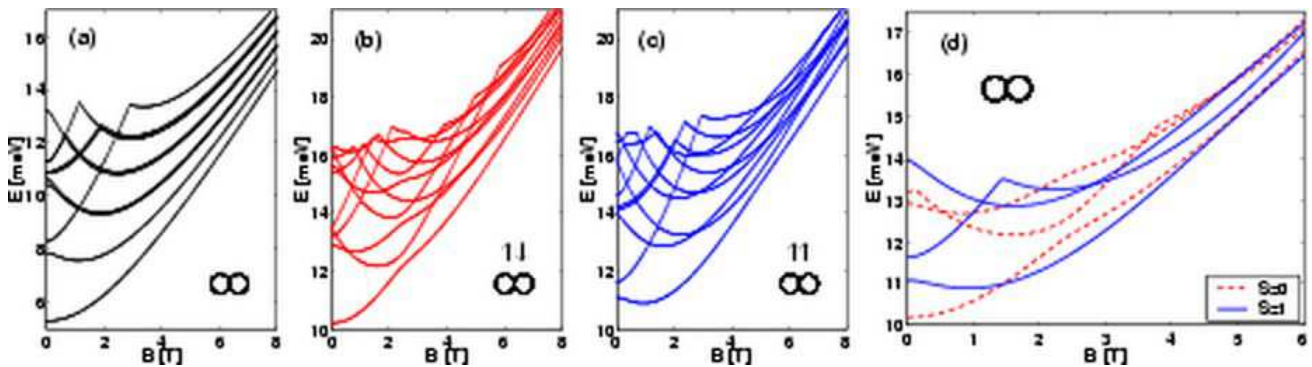} 
\caption{Energy levels of $L_x=5, L_y=0$ two-minima QDM. See
Fig.~\ref{elevLx0Ly0} for details.}
\label{elevLx5Ly0}
\end{figure*}

\begin{figure}
\includegraphics*[width=\columnwidth]{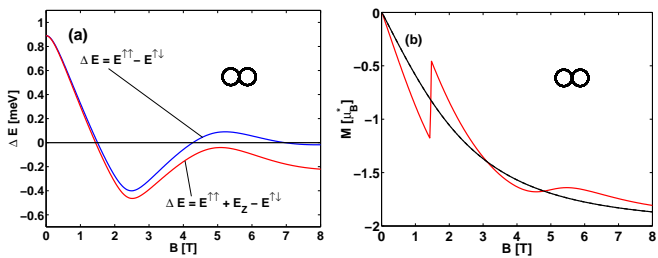}  
\caption{Triplet-singlet energy difference (a) and magnetization (b)
for $L_x=5, L_y=0$ two-minima QDM. See Fig.~\ref{dELx0Ly0} for
details.}
\label{dELx5Ly0}
\end{figure}

\subsection{Energy levels of $L_x = 5$ nm double dot}

\begin{figure}
\includegraphics*[width=\columnwidth]{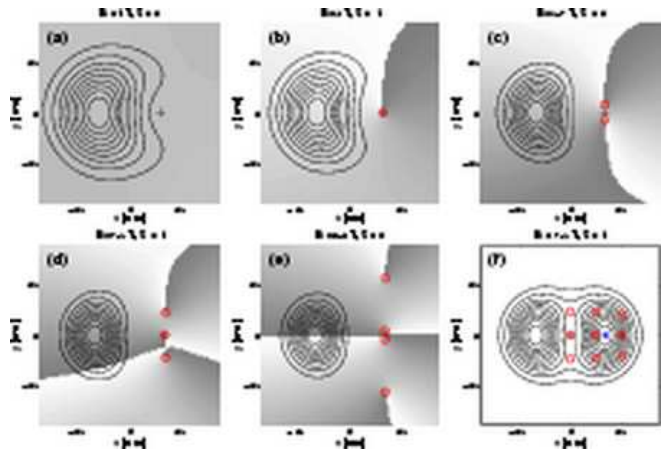}  
\caption{(a)-(e) contours of conditional densities $|\psi_c(x,y)|^2$ and phase
  of the conditional wavefunction $\theta_c(x,y)$ in gray-scale for
  $L_x=5, L_y=0$ two-minima QDM. See Fig.~\ref{vortexLx0Ly0} for
  details.  (f) contours of \emph{total} electron density of the three-vortex
  triplet state are plotted in the background and positions of the
  vortices with the fixed electron in three different positions.}
\label{vortexLx5Ly0}
\end{figure}

\begin{figure}
\includegraphics*[width=0.6\columnwidth]{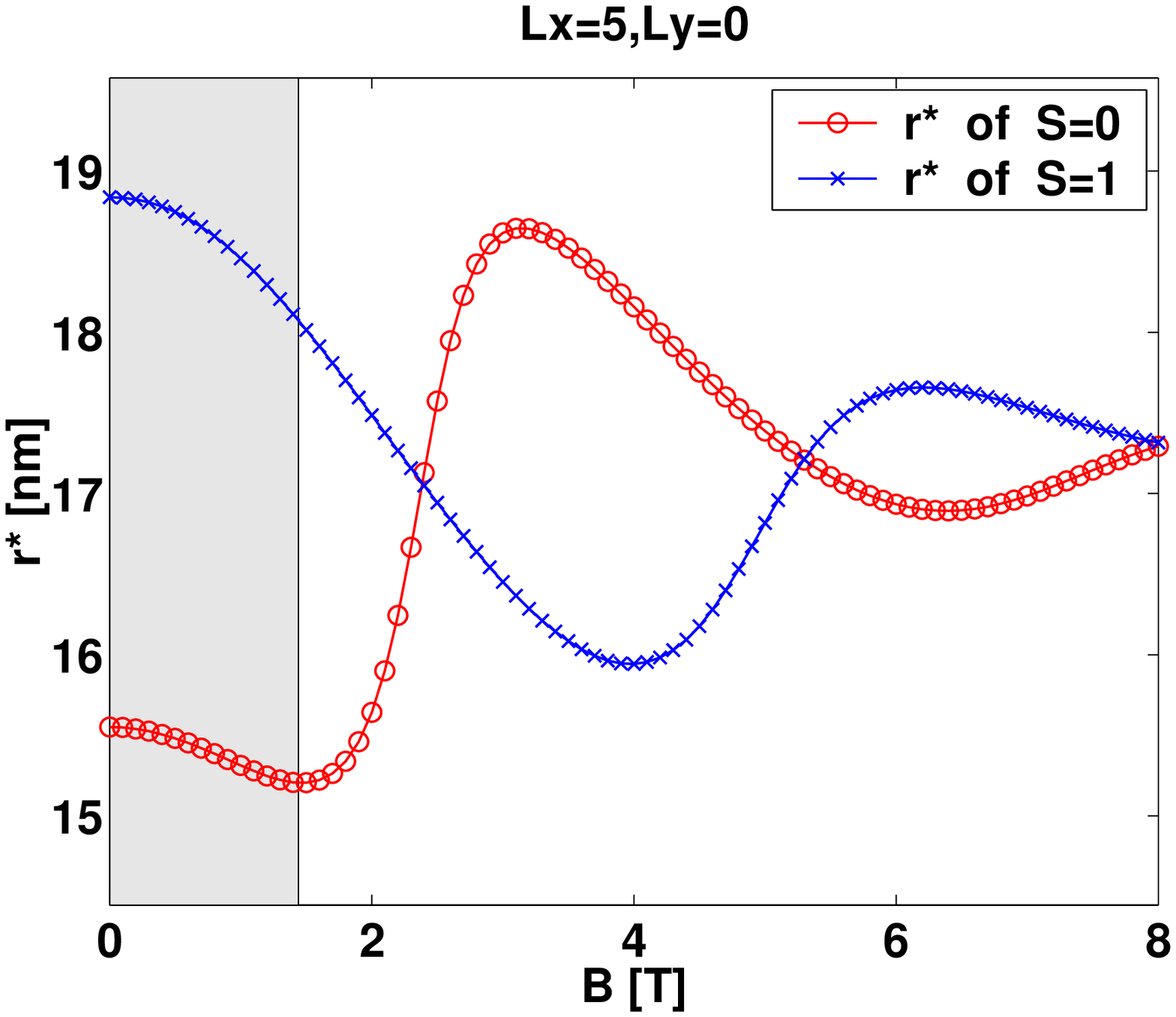}
\caption{Most probable position in nm of singlet ($S=0$) and triplet ($S=1$)
states for two-minima ($L_x=5,L_y=0$ nm) QDM. Singlet ground state
magnetic field region is marked with gray background color.}
\label{rstarLx5Ly0}
\end{figure}

\begin{figure}
\includegraphics*[width=\columnwidth]{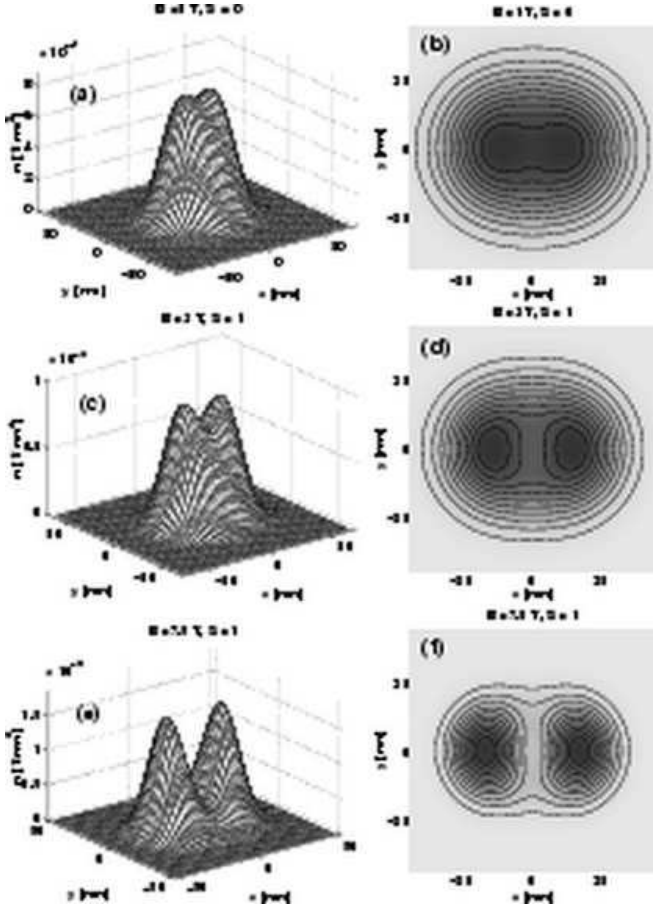}  
\caption{Density of the ground state at different magnetic field
values for two-minima ($L_x=5,L_y=0$ nm) QDM. See
Fig.~\ref{densLx0Ly0} for details.}
\label{densLx5Ly0}
\end{figure}

We choose one distance between the dots, $L_x=5$ nm, to study the
properties of the double dot in more detail. We plot energy levels,
singlet-triplet splitting, magnetization, vortices, the most probable
position and total electron density of this double dot in
Figs.~\ref{elevLx5Ly0} - ~\ref{densLx5Ly0}.  The energy levels in
Fig.~\ref{elevLx5Ly0} are now modified, compared to the parabolic QD,
due to the lower symmetry of the confinement potential. The circular
symmetry is no longer present. The lower symmetry shifts and splits
degenerate levels.  The non-interacting levels, in
Fig.~\ref{elevLx5Ly0} (a), split at zero magnetic field and there is
also a small anticrossing of levels, just barely visible in the
figure. Also degenerate levels of the single QD (see
Fig.~\ref{elevLx0Ly0} (a)) are now slightly displaced in energy, which
is mostly seen as thicker lines in Fig.~\ref{elevLx5Ly0} (a). In the
interacting spectra, (b) and (c), we see many anticrossings. Also the
nature of the lowest level does not change abruptly with crossing
levels as in the single QD, but instead we see anticrossing levels.
For example, there are clear anticrossings in the double dot singlet
states of Fig.~\ref{elevLx5Ly0} (b) whereas states cross in the
singlet state of parabolic QD of Fig.~\ref{elevLx0Ly0} (b).

We plot three lowest singlet and triplet energy levels in the same
figure to see the transition points and energy differences between the
states more clearly (Fig.~\ref{elevLx5Ly0} (d)). The ground state is a
singlet at small $B$, also in the double dot, and later it changes to
triplet. The Zeeman term lowers the triplet energy enough so that no
second singlet (ground) state is observed at higher $B$, even though
the singlet energy becomes very close to the triplet energy at $B
\approx 5$ T, as can be seen in Fig.~\ref{elevLx5Ly0} (d). The
interesting anticrossing of spin singlet between $B=2$ and $3$ T is
now an excited state as the triplet is the ground state. We also find
anticrossing ground state levels in the spin triplet around $B \approx
5.5$ T, but the repulsion of levels is not so clear at high $B$.

\subsection{Singlet-triplet splitting and magnetization of $L_x = 5$ nm double dot}

The energy difference between the lowest triplet and singlet states as
a function magnetic field in the $L_x=5$ nm double dot is plotted in
Fig.~\ref{dELx5Ly0} (a), and the magnetization in Fig.~\ref{dELx5Ly0}
(b). The sharp jump in the magnetization corresponds to the
singlet-triplet transition.  Even if the system is not circular
symmetric, and angular momentum is not a good quantum number, there is
an increase of the expectation value of angular momentum at the
transition.  We can clearly see that the magnetization increases
suddenly at the transition point. Around $B \approx 5.5$ T there is a
bump in the magnetization. This is exactly at the anticrossing point
of the triplet state. Therefore the symmetry of the triplet state
changes or the magnetic moments of the electrons change. This time it
is not seen as an abrupt change but as a continuous one. Similar
magnetization curves are seen in asymmetric QDs with a correct
deformation from the parabolic confinement to a more wire-like
confinement.\cite{Drouvelis}

\subsection{Vortices of $L_x = 5$ nm double dot}

In the case of QDMs the states cannot be identified with angular
momentum since it is not a good quantum number. However, we can still
study vortices and conditional density of the double dot. We
fix one electron at the most probable position at ${\bf
r}^*=(x^*,0)$ and study the conditional single-particle
wave function 
\begin{equation}
\psi_c(x,y) =
\frac{\Psi_S[(x,y),(x^*,0)]}{\Psi_S[(-x^*,0),(x^*,0)]}.
\end{equation} 
The most probable position of the two-minima QDM lies always on the
$x$ axis.  In Fig.~\ref{vortexLx5Ly0} (a), at low $B$, no vortices are
found and the phase is constant. Contours are again localized to the
left of the fixed electron having the maximum at ($-x^*,0$). The
conditional density in a double dot is more localized compared to the
conditional density of a single QD in Fig.~\ref{vortexLx0Ly0} (a).
The next plot, Fig.~\ref{vortexLx5Ly0} (b) shows data for the triplet
at $B=3.0$ T with one vortex and (c) shows the singlet state at
$B=5.7$ T with two vortices, (d) the triplet at $B=7.5$ T with three
vortices, and (e) the singlet at $B=8.2$ T with four vortices. The
singlet state is not a ground state after the first singlet-triplet
transition which means that two- and four-vortex solutions in (c) and
(e) are only found as excited states.

In the double dot vortices appear in the dot similarly as in the
single QD. The vortices, on the other hand, are not exactly on top of
the fixed electron, and they can be found at considerable distances
from the fixed electron, as in Fig.~\ref{vortexLx5Ly0} (e). Looking at
the ground state at $B=7.5$ T in Fig.~\ref{vortexLx5Ly0} (d) in more
detail, it is interesting to see that the vortices are seen in the
vicinity of the fixed electron even if the conditional density is
localized closer to the other dot and contour lines are rather
circular though flattened in the $x$ direction due to Coulomb
repulsion. Despite the fact that the density of an electron is rather
localized to one of the dots the two electrons move in a strongly
correlated way in the double dot. The dynamics of vortices can be
studied by changing the position of the fixed electron.  In
Fig.~\ref{vortexLx5Ly0} (f) the electron is fixed in three different
positions for the $B=7.5$ T triplet ground state. The \emph{total}
electron density is plotted in the background.  The vortices follow
the fixed electron which leads us to conclude that also in a
non-symmetric potential flux quanta and electrons form composite
particles.  It is surprising to find composite particle solutions
of electrons and flux quanta in the non-parabolic symmetry as
well.~\cite{AriPRL}

The vortices are not exactly on top of the fixed electron. This may be in part
due to the finite-size basis expansion, but on the other hand there is no
reason why they should be \emph{exactly} on top of the fixed
electron. Interesting vortex clusters of a six-electron parabolic QD are
studied in a recent article (Ref.~\onlinecite{HenriVortex}) and also in
elliptical and rectangular QDs in Ref.~\onlinecite{HenriVortex2}. In the
double dot, the vortex patterns change continuously, where always an extra
pair of vortices approach the fixed electron from minus and plus infinity from
the $y$ axis lying in the same vertical line with the fixed electron as in
Figs.~\ref{vortexLx5Ly0} (c)-(e).

\subsection{Most probable position and total electron density 
of $L_x = 5$ nm double dot}

The most probable position (${\bf r}^*$) for the lowest singlet and
triplet states is plotted in Fig.~\ref{rstarLx5Ly0}. Due to the
anticrossing levels also the singlet and triplet most probable
positions change continuously. A similar outward relaxation of the
conditional density in the double dot is associated to the approaching
vortices as was seen for the parabolic QD associated with the sudden
change in the angular momentum state.

Ground state densities at three different magnetic field values are
shown in Fig.~\ref{densLx5Ly0}. They show how the electron
density localizes into the two minima as the magnetic field
increases. The interacting density shows two peaks also in the
low-field regime whereas the non-interacting density still
has just one maximum (see the dotted line in Fig.~\ref{ddpot} (c)).

\begin{figure}
\includegraphics*[width=\columnwidth]{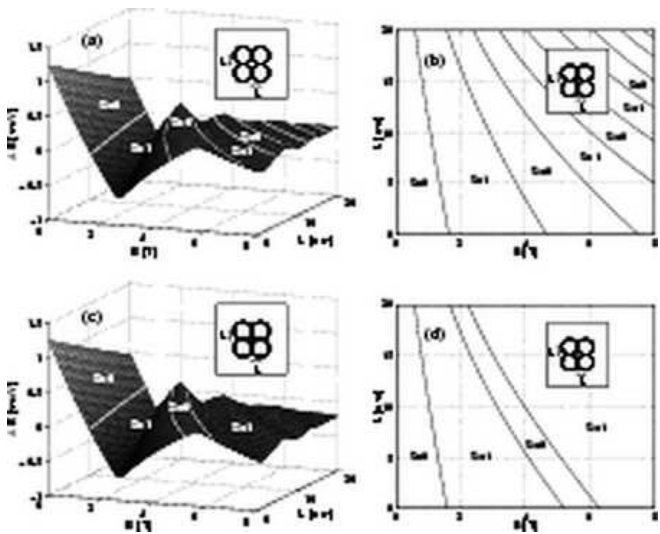}
\caption{Triplet-singlet energy difference ($\Delta E = E^{\uparrow
\uparrow} - E^{\uparrow \downarrow}$) as a function of magnetic field
in square-symmetric ($L=L_x=L_y$) four-minima quantum dot
molecule. See Fig.~\ref{ELx5Ly0} for details.}
\label{ELx5Ly5}
\end{figure}

\section{Square-symmetric four-minima quantum-dot molecule ($L_x=L_y \neq 0$)}
\label{Lx5Ly5}

In this Section we study two electrons in lateral four-minima
quantum-dot molecules. The minima are arranged in the way that they
form a square in the lateral direction.

\subsection{Singlet-triplet splitting as a function of $L$}

We study singlet and triplet states as a function of magnetic field
and dot-dot separation.  Fig.~\ref{ELx5Ly5} (a) shows altering singlet
and triplet states as a function of magnetic field. More frequent
singlet-triplet changes are seen at greater separations between the
dots (large $L$). Also notable is the large energy difference of the
second singlet state to the triplet state.  The second singlet also
persists as a ground state to the greatest studied separation $L$,
which is not true in the double dot, if the Zeeman energy is included
(Fig.~\ref{ELx5Ly5} (c)). In all studied separations $L$
the magnetic field evolution of the Zeeman coupled four-minima QDM
(Figs. (c) and (d)) follows the same pattern. At small magnetic field
values the ground state is singlet, then triplet and again singlet in
a small magnetic field window before the ground state changes to
triplet permanently. However, with large separations $L$ the system
becomes spin-polarized at lower magnetic field values, \ie the border
line of the second singlet and second triplet curves towards low-field
region with increasing $L$.

We will now analyze the rapid changes of the singlet and triplet
states of four-minima QDM.  Singlet-triplet (and triplet-singlet)
transition points shift to lower magnetic field values at greater $L$,
where the area of the QDM ($A$) is effectively larger. If the
transitions occur at effectively the same values of the magnetic flux
($\Phi=BA$), the transitions should be seen at lower $B$ when the area
is larger.  This explains why the border lines between singlet and
triplet states curve towards lower $B$ at greater separations between
the dots.

The second singlet is seen as a ground state in the Zeeman coupled
system even at the very large distance of $L=20$ nm where the
perturbation from a purely parabolic potential is clear. However, in
this type of square- or ring-like potential spatially symmetric states
(singlet) are energetically more favorable than in elongated
potentials (double dot) where in general the spatially antisymmetric
states (triplet) are favored without paying too high price in Coulomb
energy. Of course, as all energy scales are quite equal, the ground
state is a delicate balance between kinetic, confinement and Coulomb
energies as a function of magnetic field.

\begin{figure*} 
\includegraphics*[width=2\columnwidth]{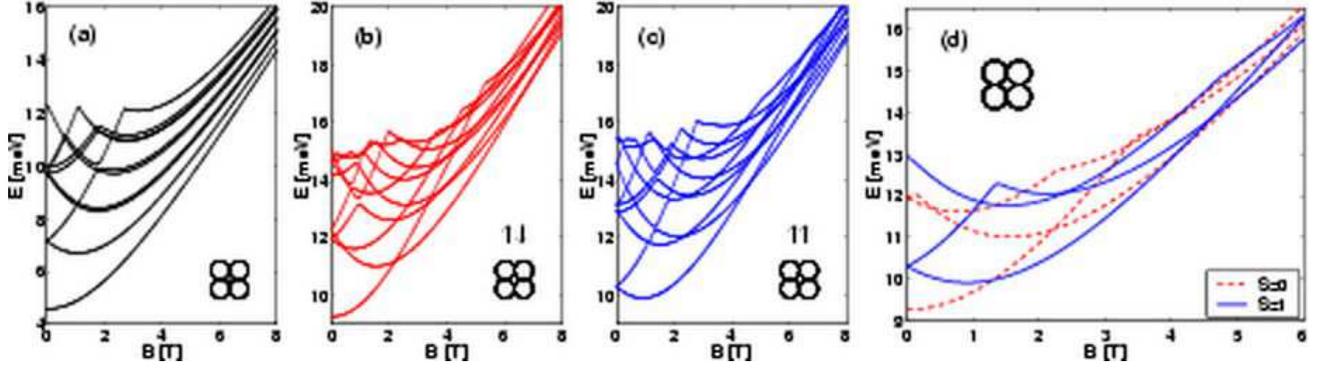}
\caption{Energy levels of $L_x=L_y=5$ nm four-minima QDM. See
Fig.~\ref{elevLx0Ly0} for details.}
\label{elevLx5Ly5}
\end{figure*}

\begin{figure} 
\includegraphics*[width=\columnwidth]{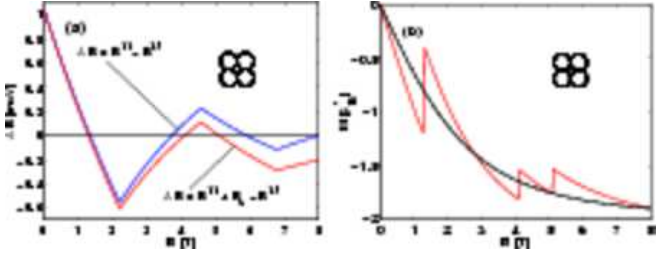}
\caption{Triplet-singlet energy difference in (a) and magnetization in
(b) for $L_x=5=L_y=5$ QDM. See Fig.~\ref{dELx0Ly0} for details.}
\label{dELx5Ly5}
\end{figure}

\begin{figure}
\includegraphics*[width=\columnwidth]{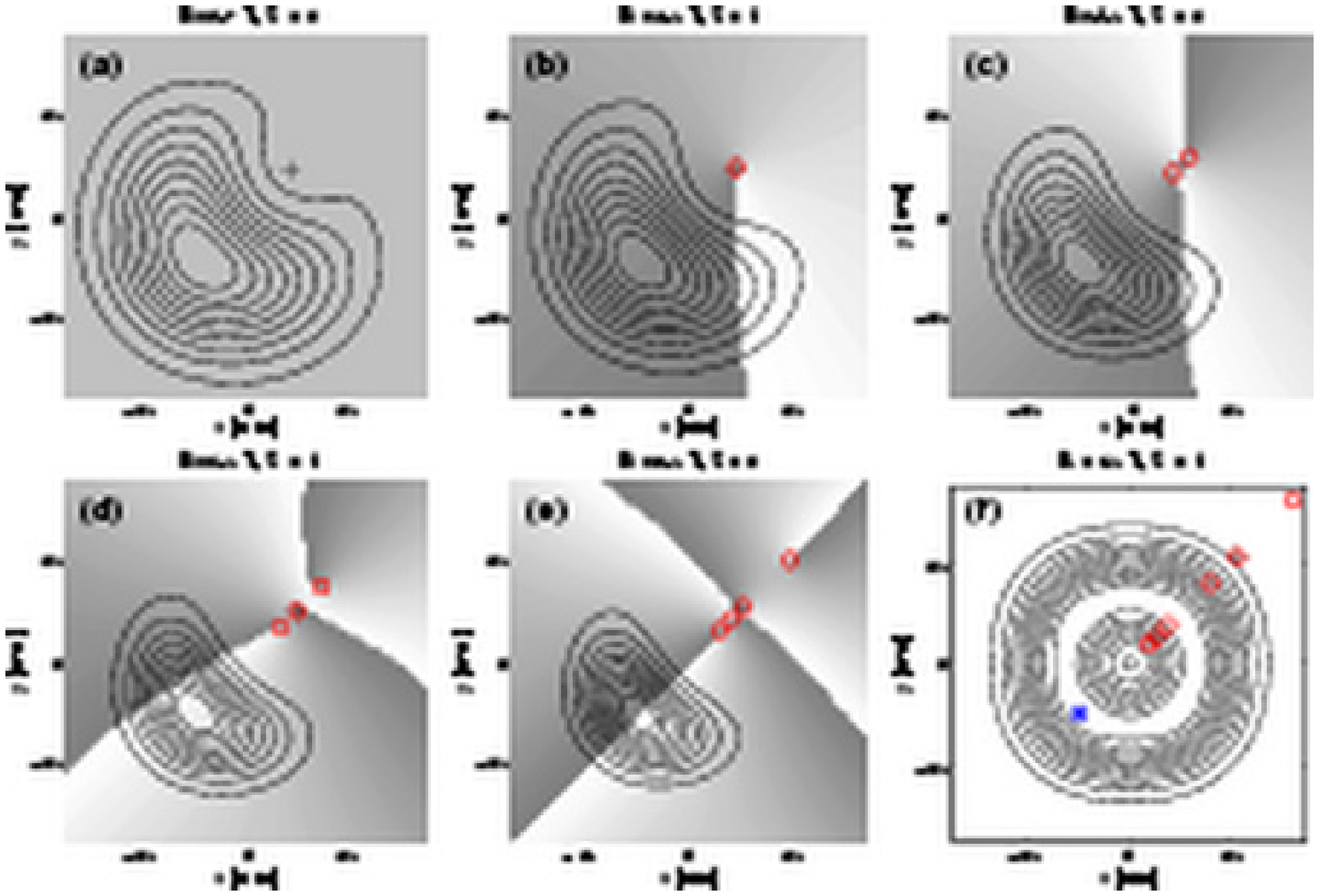}
\caption{(a)-(e) Contours of conditional densities $|\psi_c(x,y)|^2$
  and phase of the conditional wavefunction $\theta_c(x,y)$ in
  gray-scale for $L_x=5, L_y=5$ four-minima QDM. See
  Fig.~\ref{vortexLx0Ly0} for details.  (f) contours of \emph{total}
  electron density of the three-vortex triplet state are plotted in
  the background and positions of the vortices with the fixed electron
  in two different positions.}
\label{vortexLx5Ly5}
\end{figure}

\begin{figure}
\includegraphics*[width=0.6\columnwidth]{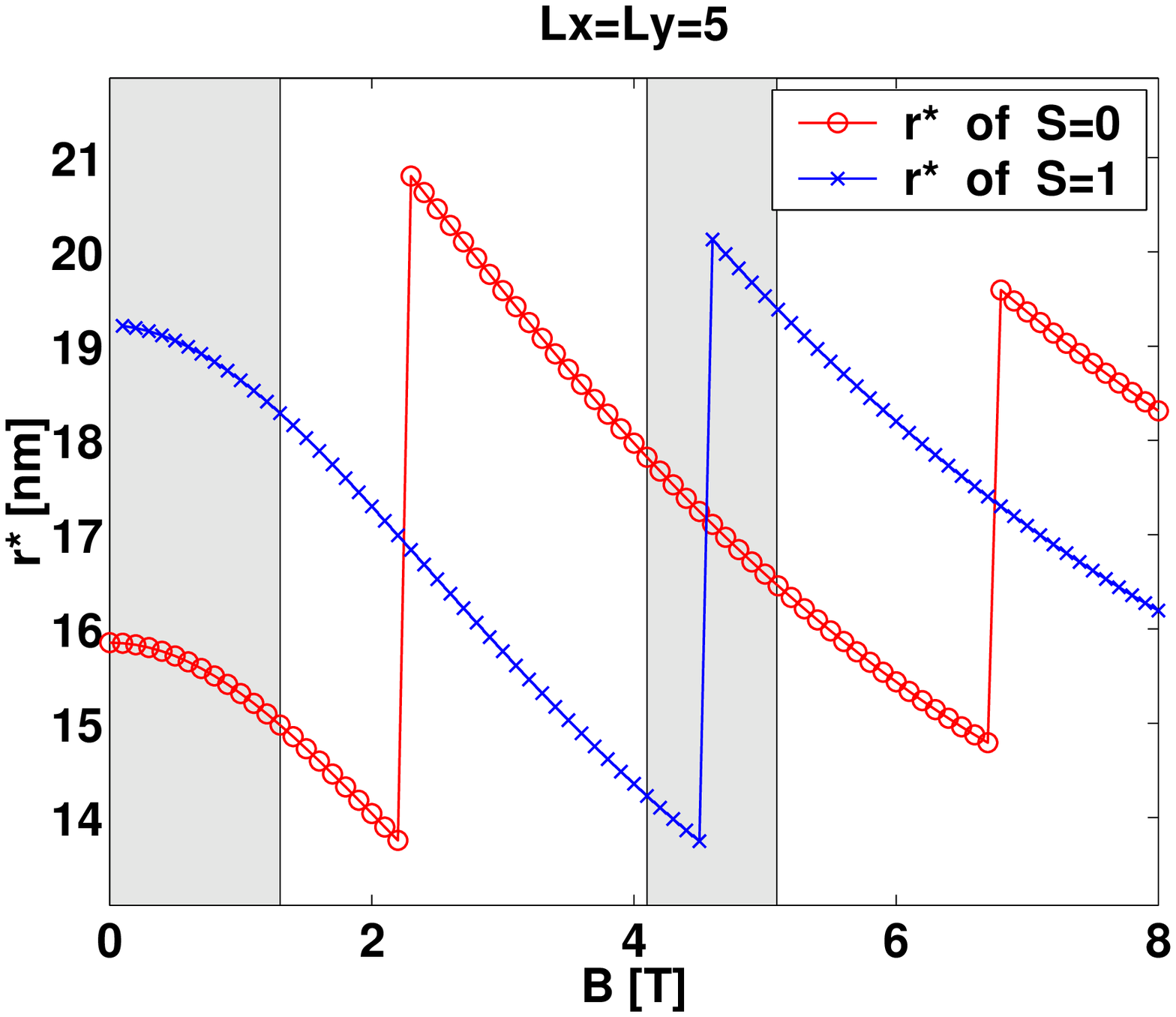}
\caption{Most probable position in nm of singlet ($S=0$) and triplet ($S=1$)
states for four-minima $L_x=L_y=5$ nm QDM. Singlet ground state
magnetic field region is marked with gray background color.}
\label{rstarLx5Ly5}
\end{figure}

\begin{figure}
\includegraphics*[width=\columnwidth]{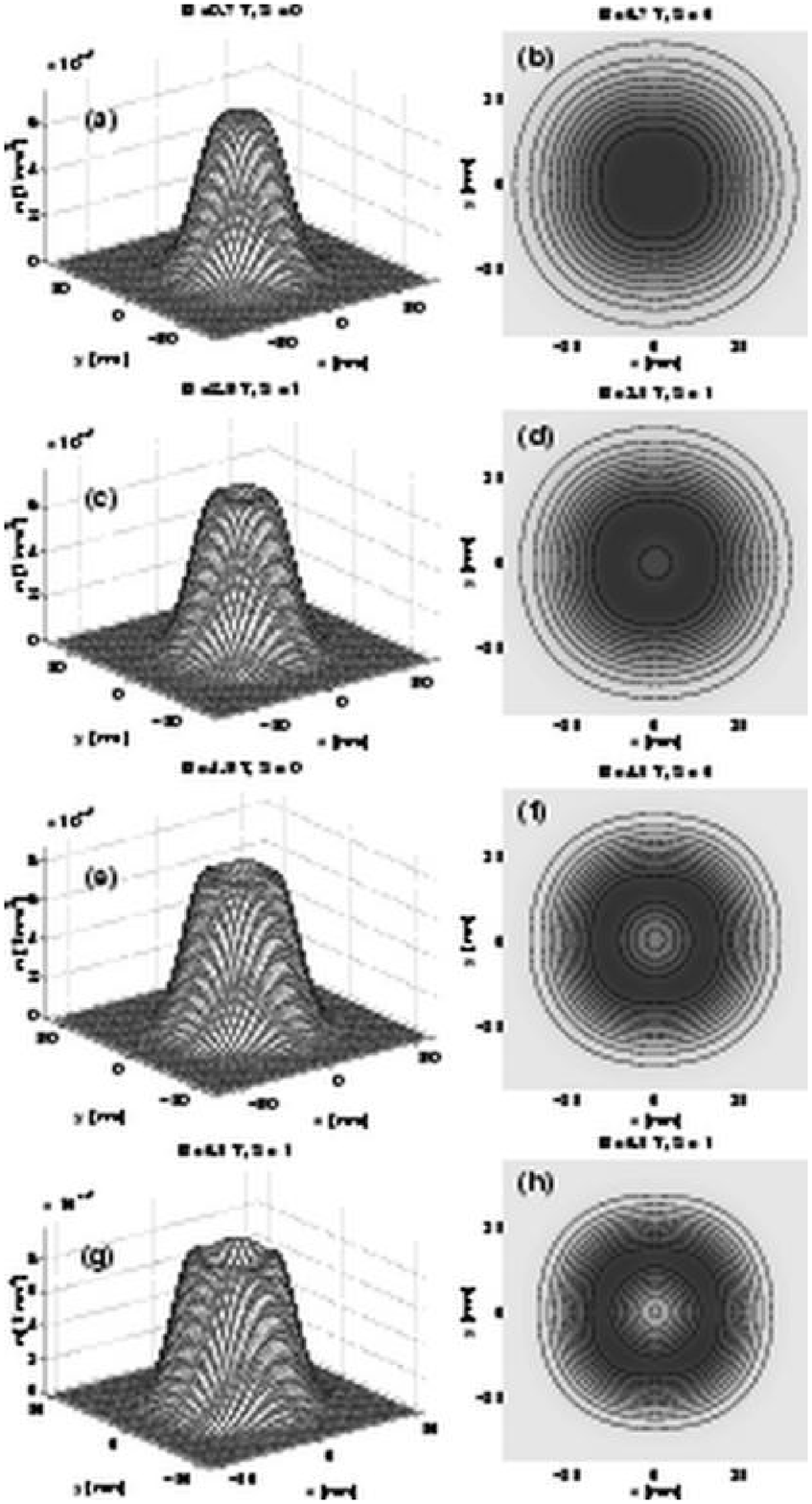}
\caption{Density of the ground state at different magnetic field
values for four-minima $L_x=L_y=5$ nm QDM. See Fig.~\ref{densLx0Ly0}
for details.}
\label{densLx5Ly5}
\end{figure}

\subsection{Energy levels of $L_x=L_y=5$ nm QDM}

We will now focus on the $L_x=L_y=5$ nm four-minima QDM. The energy
levels of non-interacting two-body, singlet and triplet states are
plotted in Fig.~\ref{elevLx5Ly5} (a)-(c), respectively. In the
square-symmetric four-minima QDM the \emph{ground state} levels do not
anticross as in the double dot, instead crossing ground
states of the same spin state are seen, as in the single QD. However,
the four-minima QDM is not circular symmetric and anticrossings are
still seen in the higher energy levels, which is not the
case with circular symmetric parabolic QD. Notice that in the
non-interacting spectra in Fig.~\ref{elevLx5Ly5} (a) many levels are
degenerate at $B=0$ for the four-minima QDM, whereas in the double dot
the zero-field degeneracies are mostly lifted, see
Fig.\ref{elevLx5Ly0} (a). Yet, many levels that are
degenerate in the single QD, Fig.\ref{elevLx0Ly0} (a), as function of
magnetic field, are slightly split in the non-interacting
energy levels of four-minima QDM.

The singlet and triplet energies can be seen in the same plot in
Fig.~\ref{elevLx5Ly5} (d). The energy levels of the four-minima QDM
in Fig~\ref{elevLx5Ly5} (d) look very similar to single QD energy
levels of Fig~\ref{elevLx0Ly0} (d). Even if the absolute values of
energies and transition points are different, the only notable
differences between parabolic QD and four-minima QDM are the small
bending at zero field of the third triplet level, near $E \approx
12.9$ meV, and small anticrossing of the uppermost triplet level at $B
\approx 1.3$ T.

\subsection{Singlet-triplet splitting and magnetization of $L_x=L_y=5$ nm QDM}

The energy difference between the lowest triplet and singlet and the
magnetization are plotted in Fig.~\ref{dELx5Ly5}. Now, as no
anticrossings of ground states are present, the triplet-singlet energy
difference shows peaks.  In the magnetization we also observe sharp
peaks whenever two ground states cross.  Similar results are seen in a
square quantum dot with a repulsive impurity of
Ref. \onlinecite{ShengXu} where the magnetization of two electrons in
the dot shows sharp transitions since no anticrossings in the ground
states are present. On the other hand, in a square QD with eight
electrons slightly rounded magnetization curves are
seen.\cite{EsaRECQD} Therefore, if the circular symmetry of the
confinement is broken, the magnetization depends on the symmetry of
the confinement but also on the number of the electrons in the QD. It
may not be straightforward to draw any conclusion about the underlying
potential from the magnetization curves. Anticrossings, on the other
hand, are clear signatures of a broken circular symmetry.

\subsection{Vortices of $L_x=L_y=5$ nm QDM}

We can identify the changes in the magnetization to the increasing
number of vortices in the two-electron QDM or to the increase of the
expectation value of relative angular momentum of the
electrons. Vortex patterns and conditional densities are shown in
Fig.~\ref{vortexLx5Ly5}. The most probable position is now found from
the line connecting two minima diagonally.
\begin{equation}
\psi_c(x,y) =
\frac{\Psi_S[(x,y),(x^*,y^*)]}{\Psi_S[(-x^*,-y^*),(x^*,y^*)]}, \ \
x^*=y^*.
\end{equation}
At $B=0.7$ T in (a) the conditional density is spread to the area of
three unoccupied dots with a peak in the furthermost dot on the
diagonal from the fixed electron. At high magnetic field the density
becomes more localized closer to the most distant minimum, in the
diagonal from the fixed electron. However, the contours show that the
conditional density is not as circularly symmetric as in a double dot,
but actually resembles more the conditional density of the single
parabolic QD. The peak in the confinement potential at the origin
seems not to affect the conditional density considerably when compared
to the single QD. At high $B$ the Coulomb repulsion forces electron
density to the outer edges of the confinement, which might not result
in very different results when compared to the single dot.  However,
the distance $L_x=L_y=5$ nm in the confinement is not particularly
large and the perturbation from the parabolic confinement is not very
large. On the other hand, altering singlet and triplet states persist
to the greatest studied distance ($L=20$ nm) between the dots of
$L=20$ nm where the perturbation from the parabolic confinement is
clear.

The vortices appear in the four-dot QDM sequentially as a function of
magnetic field. At low $B$ the ground state is a singlet with no
vortices, then a triplet with one vortex, a singlet with two vortices
and then a triplet with three vortices. The singlet with four vortices
in Fig.~\ref{vortexLx5Ly5} (e) is an excited state as the system
becomes spin polarized after the two-vortex singlet state.  The
vortices are located in the diagonal going through the fixed electron
position, see Figs.~\ref{vortexLx5Ly5} (c)-(e). The vortices seem to
be further away from the fixed electron (in the case of more than one
vortex) than in the single QD. This is also true in the double
dot. This could be identified to repulsion between the vortices but it
is difficult to assess since the length scales are different due to
different confinement strength and also the basis causes some
errors. However, with six electrons in a parabolic confinement one can
see a clear repulsion between the vortices.~\cite{HenriVortex}

Vortex dynamics of the three-vortex solution is studied by changing
the position of the fixed electron in Fig.~\ref{vortexLx5Ly5}
(f). Total electron density of the same state is plotted in the
background of Fig.~\ref{vortexLx5Ly5} (f).  Vortices are seen to
follow the fixed electron also in the four-minima QDM. However, in the
four-minima QDM the vortices are further away from the fixed electron
as the distance from the origin increases.

\subsection{The most probable position and density of $L_x=L_y=5$ nm QDM}

The most probable positions of the lowest singlet and triplet states
of the four-minima QDM (Fig.~\ref{rstarLx5Ly5}) show very similar
behavior as the single QD. Only the most probable positions are on
average roughly $2$ nm greater at all field strengths compared to the
single QD (size of the QDM is larger compared to single QD). Otherwise
continuously decreasing ${\bf r}^*$ shows a jump when the lowest
singlet (or triplet) state changes.

Ground state densities and contours are plotted in
Fig.~\ref{densLx5Ly5}. Starting from a rather flat density at low
fields, a hole begins to form in the center as the magnetic field is
increasing. The electron density localizes into a narrowing ring
around the origin. However, compared to the parabolic QD, there are
peaks forming in the four corners of the density, instead of a smooth
density ring. Also the density looks more square-like in all of the
contours.

\begin{figure}
\includegraphics*[width=\columnwidth]{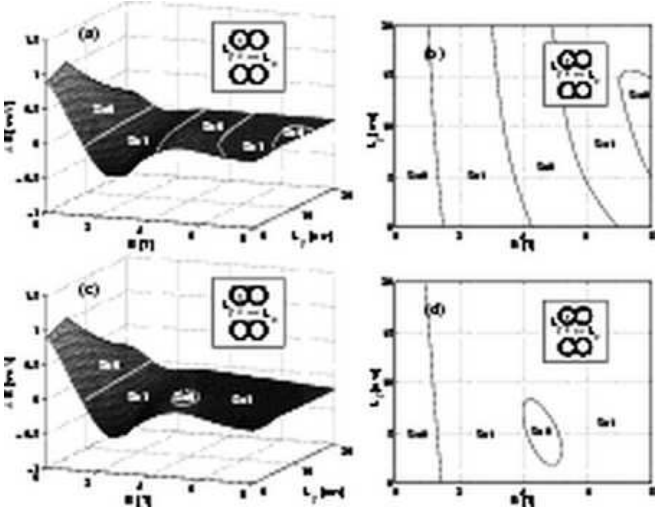}
\caption{
Triplet-singlet energy difference ($\Delta E = E^{\uparrow
\uparrow} - E^{\uparrow \downarrow}$) as a function of magnetic field
in rectangular-symmetric four-minima quantum dot molecule. $L_x$
is fixed to $5$ nm and $L_y$ is varied from $0$ to $20$ nm. Therefore
at $L_y=0$ we have $L_x=5$ nm double dot and at $L_y=5$ nm it is the
square-symmetric four-minima QDM. The energy difference is plotted as
a function of $L_y$ and magnetic field in (a) without
Zeeman energy and in (c) with the Zeeman energy included ($\Delta E =
E^{\uparrow \uparrow} + E_Z - E^{\uparrow \downarrow}$). In (b) and
(d) the ground state regions of the singlet and triplet states are
plotted as function of $B$ and $L_y$, without and with Zeeman energy,
respectively.}
\label{ELx5Ly10}
\end{figure}

\section{Rectangular four-minima quantum dot molecule 
($L_x \neq L_y \neq 0$) }
\label{Lx5Ly10}

In this Section we examine the triplet-singlet energy difference,
energy eigenlevels, magnetization, vortices, the most probable
positions, and densities of four-minima QDM with
rectangular positioning of the QD minima in the lateral plane.

\subsection{Singlet triplet splitting as function of $L_y$ for fixed 
$L_x=5$ nm rectangular QDM}

\begin{figure*} 
\includegraphics*[width=2\columnwidth]{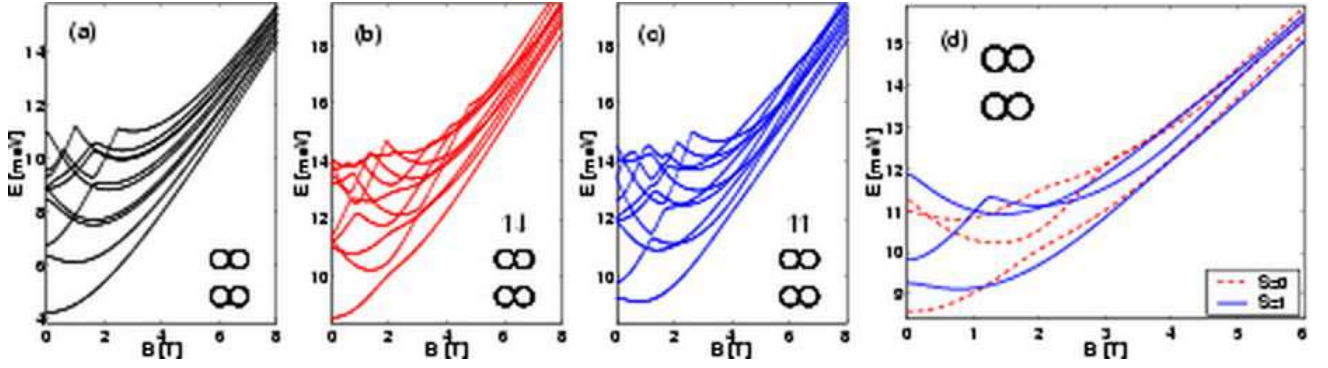}
\caption{Energy levels of $L_x=5, L_y=10$ nm rectangular
four-minima QDM. See Fig.~\ref{elevLx0Ly0} for details.}
\label{elevLx5Ly10}
\end{figure*}

The triplet-singlet energy difference of the rectangular QDM is plotted in
Fig.~\ref{ELx5Ly10}. The distance between the minima in the $x$
direction is fixed while the distance in the $y$ direction is varied.
We set $L_x=5$ nm and vary $L_y$ from zero to $20$ nm. Therefore, at
$L_y=0$ we have a $L_x=5$ nm double dot, and at $L_y=5$ nm we have the
$L_x=L_y=5$ nm square-symmetric four-minima QDM studied in the
preceding section.  The smooth surface in Fig.~\ref{ELx5Ly10} is due
to anticrossing ground states similarly as in the double dot. The
anticrossings in the lowest levels of the singlet and triplet states
are again present if the symmetry is distorted from a square to a
rectangular symmetry, see the energy levels in
Fig.~\ref{elevLx5Ly10}. The only sharp peaks in Fig.~\ref{ELx5Ly10}
(a) and (c) correspond to rectangular symmetric QDM at $L_x=L_y=5$ nm.

One can also see that as a function of magnetic field singlet and
triplet states do not change as rapidly as in the square-symmetric
four-minima QDM. Interestingly, the third singlet region terminates
around $L_y \approx 15$ nm. So at sufficiently large distance between
the two $L_x=5$ nm double dots the singlet state is no longer
favorable even if the Zeeman term is excluded. In the case of a double
dot in Ref.~\onlinecite{AriPRL} it was not possible to say whether the
second singlet state would terminate at greater distances between the
dots, but for two double dots the third singlet region clearly
terminates.

\begin{figure} 
\includegraphics*[width=\columnwidth]{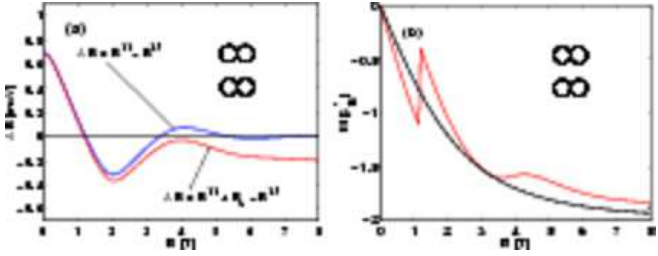}
\caption{Triplet-singlet energy difference (a) and magnetization (b)
in $L_x=5, L_y=10$ nm rectangular four-minima QDM. Magnetization is
given in the units of effective Bohr magnetons $\mu_B^* = e \hbar/2
m^*$.}
\label{dELx5Ly10}
\end{figure}

If the Zeeman term is included (Figs.~\ref{ELx5Ly10} (c) and (d)) the
second singlet state can only be observed in a small region where the
rectangular four-minima QDM is close to square symmetry (near $L_y=5$
nm). Actually the energy difference has its maximum, as function of
$L_y$, at the square symmetry. If we follow the energy difference
at zero magnetic field as a function of $L_y$, it
first increases reaching the maximum at $L_y=5$ nm and then it
decreases again when $L_y$ is increased.

The stability of the singlet states (and also triplet states) in the
\emph{square}-symmetric QDM can be understood from the relatively high
energy of the triplet state (or singlet) near the peak in $\Delta E$.
In the square symmetry the degenerate energy levels at the crossing
point are energetically very unfavorable. In rectangular symmetry
degeneracies are lifted (anticrossings), which lowers the energy of
the other spin type and also reduces the energy difference, $\Delta
E$. Thus the energy differences are always smaller in the rectangular
symmetry when anticrossings are present. The Jahn-Teller theorem
states that any non-linear molecular system in a degenerate electronic
state will be unstable and will undergo a distortion to form a system
of lower symmetry and lower energy, thereby removing the
degeneracy.~\cite{JahnTeller} In a QDM, the system cannot, of course,
lower the symmetry of the external confinement spontaneously to lift
the degeneracies, but the large triplet-singlet energy differences in
the square-symmetric QDM can be understood via Jahn-Teller effect: If
the symmetry is lowered, degeneracies are lifted and smaller
triplet-singlet energy differences are observed.

\begin{figure}
\includegraphics*[width=\columnwidth]{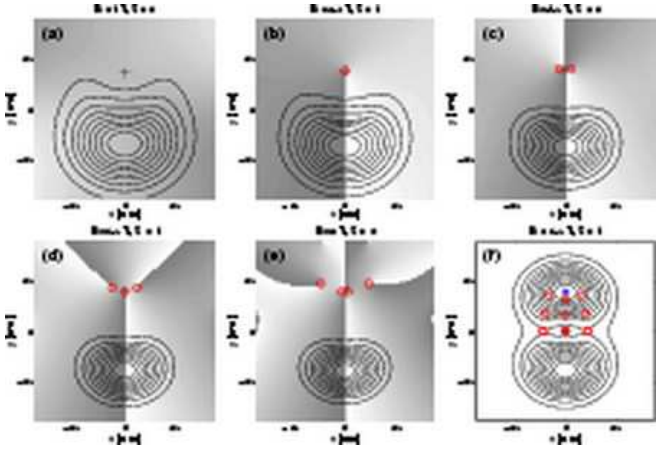}
\caption{(a)-(e) contours of conditional densities $|\psi_c(x,y)|^2$ and phase
  of the conditional wavefunction $\theta_c(x,y)$ in gray-scale for
  $L_x=5, L_y=10$ four-minima QDM. See Fig.~\ref{vortexLx0Ly0} for
  details.
  (f) contours of \emph{total} electron density of the three-vortex
  triplet state are plotted in the background and positions of the
  vortices with the fixed electron in three different positions.} 
\label{vortexLx5Ly10}
\end{figure}

\begin{figure}
\includegraphics*[width=\columnwidth]{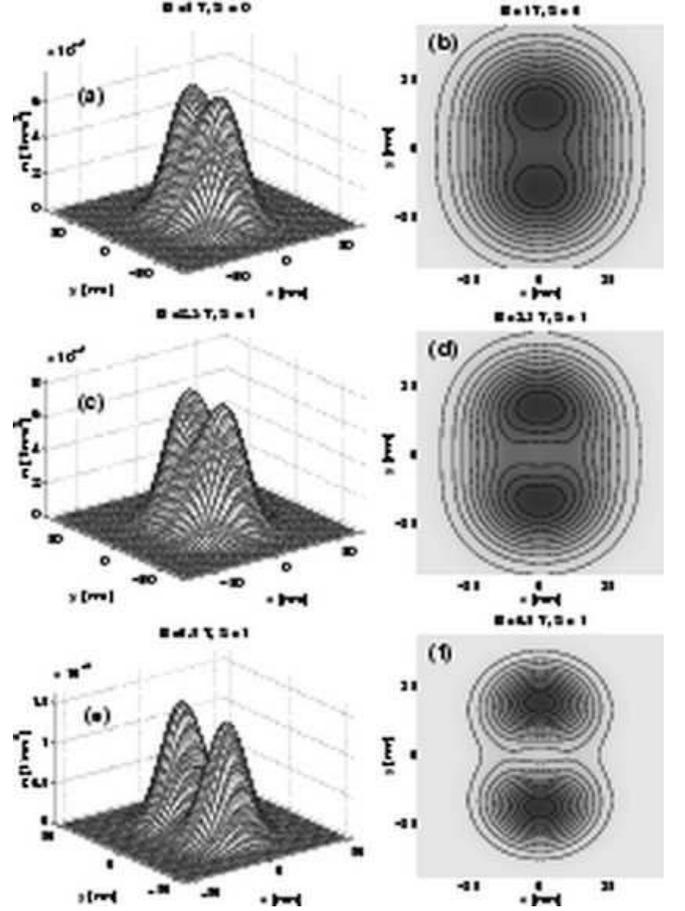}
\caption{Density of the ground state at different magnetic field
values for $L_x=5,L_y=10$ nm rectangular four-minima  QDM. See Fig.~\ref{densLx0Ly0} 
for details.}
\label{densLx5Ly10}
\end{figure}

\begin{figure}
\includegraphics*[width=0.6\columnwidth]{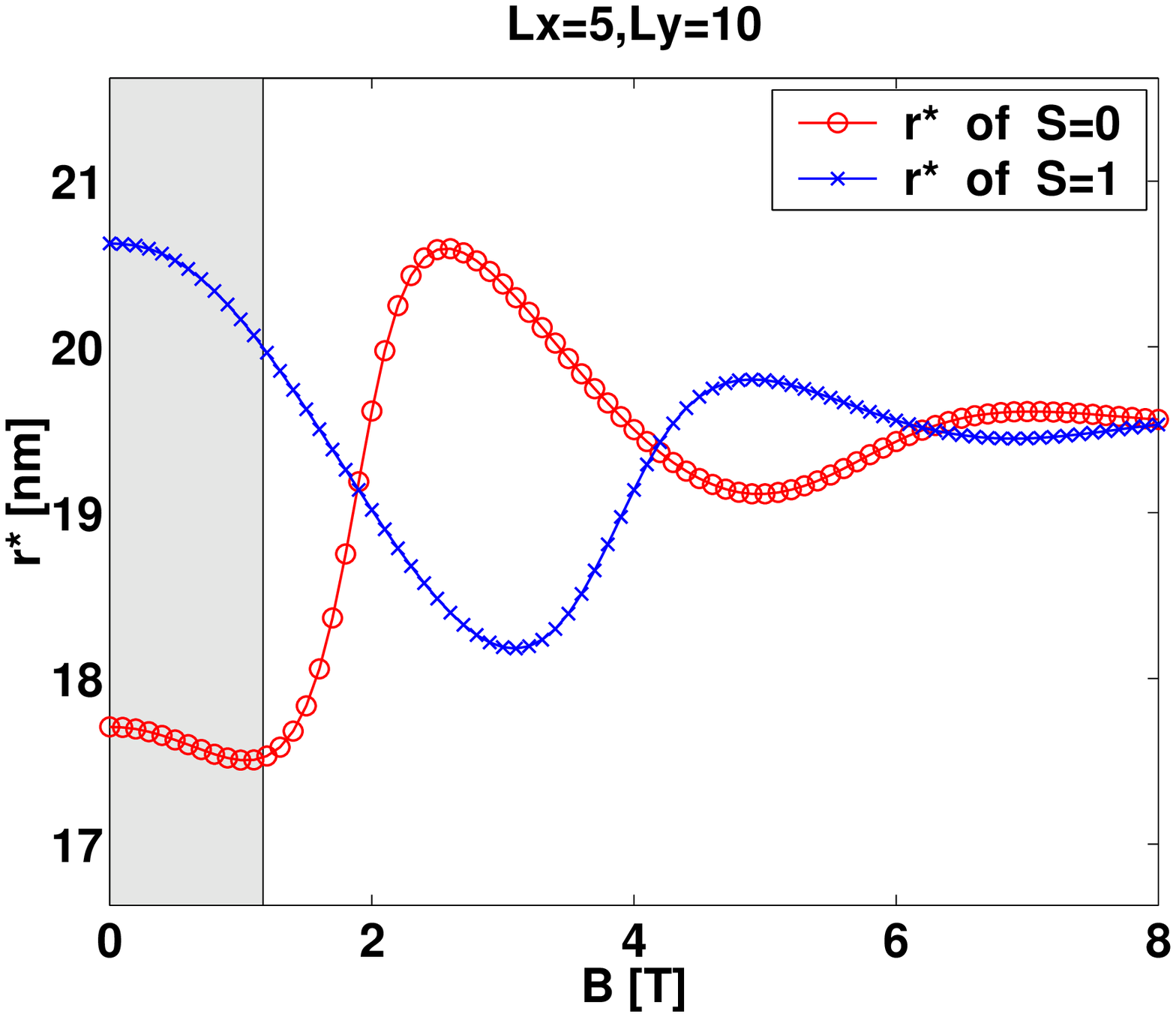}
\caption{Most probable position in nm of singlet ($S=0$) and triplet
($S=1$) states for rectangular-symmetric four-minima ($L_x=5,L_y=10$
nm) QDM. Singlet ground state (magnetic field) region is marked with
gray background color.}
\label{rstarLx5Ly10}
\end{figure}

\subsection{Energy levels of $L_x=5,L_y=10$ nm QDM}

We will now study rectangular $L_x=5,L_y=10$ nm QDM in more detail.
Fig.~\ref{elevLx5Ly10} (b) and (c) reveal many anticrossings in the
interacting two-body spectra of $L_x=5,L_y=10$ nm QDM, both in the
ground states and also in the excited states. Many features look
similar as in the double dot spectra in Fig.~\ref{elevLx5Ly0} but the
anticrossing gaps are bigger here due to the greater separations
between the dots (greater deviation from the circular
symmetry). Fig.~\ref{elevLx5Ly10} (d) shows singlet and triplet energy
levels in the same plot.  The second singlet becomes very close to the
triplet near $B=5$ T, but the triplet remains the ground state.

\subsection{Singlet-triplet splitting and magnetization of $L_x=5,L_y=10$ nm QDM}

The energy difference of triplet and singlet states is plotted in
Fig.~\ref{dELx5Ly10} (a). The magnetization in Fig.~\ref{dELx5Ly10}
(b) show first a sharp peak which corresponds to singlet triplet
transition. The next change is from the one-vortex triplet to the
three-vortex triplet and as these two states anti-cross we see
continuous increase of the magnetization before it starts to decrease
again due to the contraction of the electron density. It is
interesting that after the bump the interacting magnetization has very
similar dependence on the magnetic field as the non-interacting
magnetization.  At high enough magnetic field the electrons are
localized into individual double dots and have single-particle
properties and the spatial extents in the interacting and
non-interacting systems are not very different (see
Eq. (\ref{extent})). However, the electrons may move in a correlated
way even if they are localized into one of the double dots. One should
remember that there is also the paramagnetic part in the
magnetization, but this is constant for a given state, if the the
angular momentum is a good quantum number, and does not depend on the
magnetic field. Of course, in a non-circular symmetry the paramagnetic
magnetization may not be constant as a function of magnetic field.

\subsection{Vortices of $L_x=5,L_y=10$ nm QDM}

The phase information and the conditional densities of the rectangular
four-minima QDM are shown in Fig.~\ref{vortexLx5Ly10}. The most
probable position lies now on the $y$ axis. Another possibility would be
to have the most probable position on a line connecting the two minima
diagonally. However, as the other double dot is left with just one
electron and the distance of $L_x=5$ nm is so small that the
single-particle density is not localized to the minima of the double
dot. So the most probable position is in the $y$ axis. In rectangular
symmetry correlations force the one electron to one of the double
dots. With small $L_x$'s conditional density shows a peak
at $x=0$ as in the non-interacting two-body density in
Fig.~\ref{ddpot}.

The conditional density becomes more localized as the
magnetic field increases. The vortices appear sequentially in the QDM. The
second and third singlet states (in Fig.~\ref{vortexLx5Ly10} (c) and
(e)) are not ground states as the system becomes spin-polarized after
the first singlet-triplet transition. There is again a repulsion
between the vortices. Interesting is to note that in (e) the two more
distant vortices are positioned much closer to the fixed electron when
compared to the double dot of Fig.~\ref{vortexLx5Ly0} (e). The white
and dark regions near the borders of Fig.~\ref{vortexLx5Ly10} (e) show
the shades of phase boundaries of more distant vortices (not visible
in the figure). 

Vortex dynamics is studied in Fig.~\ref{vortexLx5Ly10} (f) for the
three vortex triplet (at $B=6.5$ T). The electron is fixed at three
different positions and the \emph{total} electron density of the same state
is plotted in the background. One vortex, or Pauli vortex, is always on
top of the fixed electron and the two additional vortices are
symmetrically on the sides of the fixed electron. As the fixed electron
is moved from the origin to the direction of the positive $y$ axis, the
vortices aside become closer to the fixed electron. 

\subsection{Total electron density and the most probable 
positions of $L_x=5,L_y=10$ nm QDM}

Ground state electron densities in Fig.~\ref{densLx5Ly10} show a
localization into two double dots as the magnetic field is
increased. If the densities would be rotated by $90$ degrees they would
resemble very much two-minima QDM (double dot) densities of
Fig.~\ref{densLx5Ly0}. The smaller displacement ($L_x=5$ nm) in the
four-minima QDM potential has a much smaller effect than the larger
displacement ($L_y=10$ nm) because electrons localize into the double
dots (with $L_x=5$ nm) separated from each other with a distance $d=2
L_y = 20$ nm. Therefore electron density of rectangular four-minima
QDM effectively resembles of that of a two-minima QDM (double
dot). This is true for interacting two-electron system.

The most probable positions of the lowest singlet and triplet states
are shown in Fig.~\ref{rstarLx5Ly10}. Continuously changing ${\bf
r}^*$ (\ie no jumps) is due to anticrossing states where symmetry of a
state (and also ${\bf r}^*$) changes continuously. Interesting is the
strong suppression of the oscillations of ${\bf r}^*$ at greater
$B$. At large magnetic field the electrons localize into a distant
double dots and interaction effects (like changing angular momentum
states in parabolic QD) have a smaller impact on the properties of the
two-electron system. The effect is quite different for parabolic QD
and square-symmetric four-minima QDM where the localization of electron
density is not so strong due to the nature of the confinement
potential.

\section{Analysis of the results and their relevance to experiments}
\label{anaa}

\subsection{Role of symmetry in quantum dot confinement}

\begin{figure}
\includegraphics[width=\columnwidth]{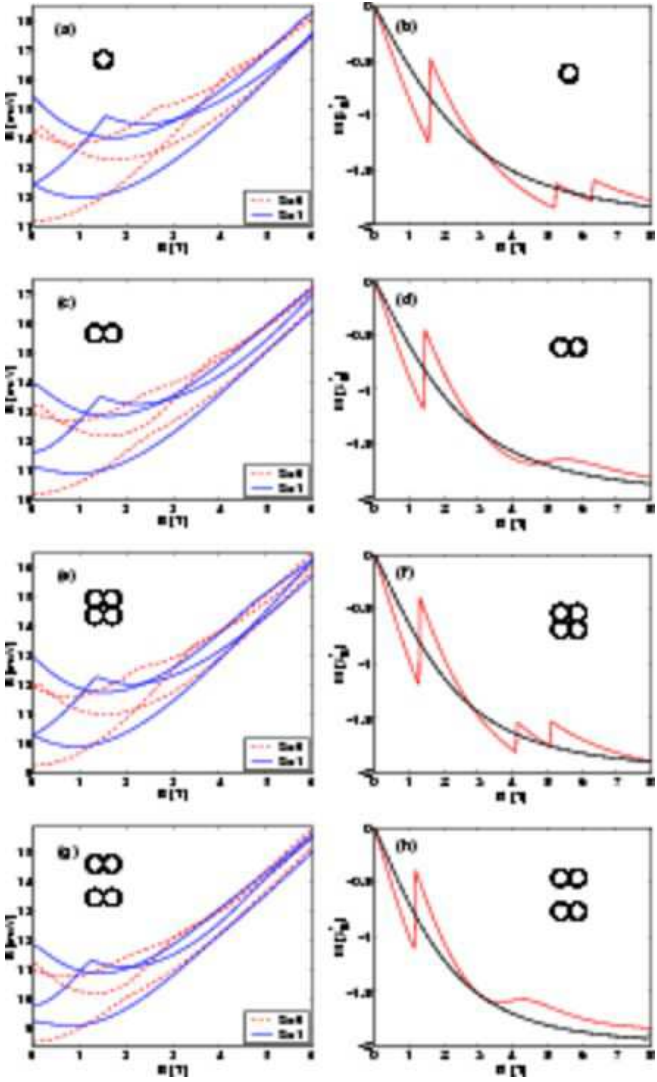}
\caption{Three lowest singlet and triplet energy levels
and magnetization for all studied quantum dot confinements.}
\label{Mess}
\end{figure}

We start our analysis of the data presented above from the measurable
quantities. One such observable is the total energy for ground and
excited states as a function of the magnetic field, as well as the
magnetization. To ease the comparison, these are collected in
Fig.~\ref{Mess}. 
The most striking feature is that the data of the the square-symmetric
four-minima QDM is very similar to the one of the single parabolic
dot. On the other hand, the data of the double dot resemble the one of
a rectangular four-minima QDM. The reason behind the similarities of
these pairs is the symmetry. The square-symmetric and circularly
symmetric cases have higher symmetries than the rectangular ones. One
can study this in detail by splitting the total Hamiltonian of
Eq.~(\ref{ham}) to two parts as $H=H_0+H_I$, where $H_0$ is the
Hamiltonian of the parabolic case ($L_x=L_y=0$), and the impurity
Hamiltonian $H_I$ contains the terms from finite $L_x$ and $L_y$
values, see Eq.~(\ref{Vc_auki}). If we now have a high symmetry in the
system, meaning $L_x=L_y$, one can see that the $H_I$ does not couple
the eigenstates of $H_0$ that have a different symmetry. On the other
hand, in the case with $L_x\neq L_y$, $H_I$ has a lower symmetry and
more of the symmetries of $H_0$ are broken. Due to this, states with
different symmetry are coupled. This leads to anticrossings in the
energies as seen in Fig.~\ref{Mess}, where also the crossings of the
high-symmetry cases are seen. One can estimate the strength of the
symmetry-lowering part from the anti-crossing gap in energy, as in the
point where the energies would cross, one has in the first
approximation a Hamiltonian matrix:
\begin{equation}
H=\left(
\begin{array}{cc}
E_0 & E_\delta\\
E_\delta & E_0 
\end{array} \right) \ ,
\end{equation}
where $E_0$ is the energy at middle of the gap, and $2 E_\delta$ is
the width of the gap.

The magnetization curve for the low- and high-symmetry cases are also
very different, see Fig.~\ref{Mess}. A common feature in all these
cases is the sharp increase in magnetization at the point where the
total spin of the ground-state changes. On the other hand, the
low-symmetry anticrossings of the energy result smooth changes in
magnetization, whereas the high-symmetry data show sudden jumps.

These findings indicate that it is rather difficult to obtain detailed
information of the system based on the energetics and the
magnetization. The symmetry of the system can be extracted, but not
much beyond that. For larger particle numbers, more and more of the
correlation effects can be captured by the mean-field level. The
effective potential has, due to the Hartree potential, a higher
symmetry than the mere external potential.\cite{MeriPRL} This results
in less details to both energetics and the magnetization.

A similar role of the symmetry can be seen on the non-measurable
quantities, like the densities. On the other hand, the vortices are
more delicate. This is because they depend linearly on the wave
function, unlike densities and energies that are second order.

\subsection{Exchange of two spins} 

The idea of double dot spin-swap operations in quantum
computing~\cite{LossDiVincenzo} lies in the coherent rotation of two
initially isolated spins. Starting with, say, spin-up electron in the
left and spin-down electron in the right dot: $|{\uparrow}\rangle_L
|{\downarrow} \rangle_R$ and rotating spins to opposite order:
$|{\downarrow}\rangle_L |{\uparrow} \rangle_R$. These would be initial
and final states of the system. Coherent rotation between initial and
final states requires entangled spin states like spin-singlet $|S
\rangle = (|{\uparrow}\rangle_L |{\downarrow} \rangle_R -
|{\downarrow}\rangle_L |{\uparrow} \rangle_R)/\sqrt{2}$ and triplet
$|T_0 \rangle = (|{\uparrow}\rangle_L |{\downarrow} \rangle_R +
|{\downarrow}\rangle_L |{\uparrow} \rangle_R)/\sqrt{2}$ whereas the
other two triplet states $|T_+ \rangle = |{\uparrow}\rangle_L |
{\uparrow} \rangle_R$ and $|T_- \rangle = |{\downarrow} \rangle_L
|{\downarrow} \rangle_R $ are not conceivable as they have identical
spins. The states described above are only for the spin-part of the
wave function but of course the spatial part, discussed extensively in
this paper for singlet and triplet eigenstates, must be modified along
with the rotation.

In the simplified Heisenberg picture, rotation depends only on the
singlet-triplet splitting energy, $J= \Delta E = E^{\uparrow
\uparrow}- E^{\uparrow \downarrow}$, or exchange coupling of two
spins.~\cite{LossDiVincenzo} Hubbard-type models can be used to study
time evolution of a little bit more elaborate
states.~\cite{Burkard_doubledot,SchliemannLossMacDonald} To fully
investigate the coherent rotation of two-electron system, it would be
better to start with initially separated electrons (that can be
constructed from many-body wave functions) and study the time
evolution of the state in the exact many-body basis instead of using
some simplified models. However, tuning $J$ with dot-dot separation at
small magnetic fields, the spin rotations can be quite safely modelled
within the Heisenberg picture.  At high magnetic fields, on the other
hand, electrons in lateral double dots form finite quantum Hall-like
composite-particle states of electrons and flux quanta, as shown in
this study and in previous studies.~\cite{AriPRL,ScarolaDasSarma}
Therefore, the coherent spin rotations at high $B$ may be quite
different from zero $B$ rotations, even if $J$ could have exactly same
value for high-$B$ and zero-$B$ states. These states are of course of
great scientific interest as their own, but from the perspective of
coherent two-spin rotations they may function quite differently as
suggested by the Heisenberg or Hubbard models. Electric control of $J$
with dot-dot separation may also be advantageous in other perspectives
because magnetic fields are more difficult to apply locally.

\subsection{Comparisons to experiments}

In very recent experiments Petta {\it et al.} demonstrate a coherent
rotation of two spins between singlet $| S \rangle$ and triplet $|T_0
\rangle$ states in a lateral double dot device.~\cite{PettaSciExp05}
They start the operation from singlet state in single QD and then
isolate two opposite spins in separated dots where the singlet-triplet
splitting vanishes and no tunneling is allowed between the two
dots. Coherent rotation is performed by bringing the two dots closer
allowing small but finite energy splitting $J$ between $|S \rangle$
and $|T_0 \rangle$ states.  Probability of finding singlet is measured
as a function of gate operation time. Fig.~\ref{ELx5Ly0} (a) of this
study shows calculated singlet-triplet splitting as a function of
dot-dot separation and magnetic field.  Following the zero magnetic
field line one and see how the singlet-triplet splitting decreases as
a function of increasing dot-dot separation. Fixing $B=0.1$ T, as in
the experiment, for single dot ($L=0$) we have $J=1.16$ meV and for
double dot at the greatest interdot distance studied ($d=2L=40$ nm) we
have $J=0.16$ meV. Petta {\it et al.}~\cite{PettaSciExp05} find three
times smaller value in single QD ($J=0.4$ meV) as our
calculations. This is simply a consequence of different quantum dot
confinement energy $\hbar \omega_0$.

Lee \etal studied experimentally singlet-triplet splitting as a
function of magnetic field in silicon two-electron double
dot.\cite{Lee} The measurements show very similar data as our
results. The first singlet-triplet transition is resolved clearly in
the experiment in accordance to our calculations.  Decreasing the
coupling between the dots results in a small shift of $J=0$ to low
fields as in our results represented in Fig.~\ref{ELx5Ly0}. The
high-field regime, where we would expect to find small $J$ and even
possibly a positive $J$, which would correspond second singlet ground
state, is not measured to very high field strengths. However, to fully
compare our results to the measurements on silicon double dots we
should recalculate our data with silicon effective mass and dielectric
constant.

Brodsky \etal~\cite{Brodsky} were the first to measure ground state
energy levels of a lateral two-electron double dot as a function of
magnetic field. Even if they did not concentrate on two-electron case
particularly, the line for two electrons is clearly visible in their
data showing also a kink indicating the singlet-triplet transition.

Magnetization is very difficult to measure directly for just two
electrons. However, indirect methods and direct methods with large
arrays of individual few-electron dots may provide interesting
experimental results.~\cite{Oosterkamp,Schwartz} Even if it is
difficult to compare existing measurements of many electron dots to
our two-electron system, the double dot measurements of Oosterkamp
\etal~\cite{Oosterkamp} show similar type of anticrossings as our
calculated magnetization curves.

\section{Summary}
\label{Summary}

In summary, we have thoroughly studied different lateral two-electron
quantum-dot molecules. All our exact diagonalization calculations were
performed for closely coupled quantum dots. Many-body electron wave
functions were allowed to extend over the whole system. We have
analyzed how the physical properties change when a deviation or
disorder is introduced in the confinement potential of the symmetric
quantum dot.  We have calculated measurable quantities such as energy
levels, singlet-triplet splitting, and magnetization as a function of
magnetic field strength. The measurable quantities were further
analyzed by calculating non-measurable quantities such as phase
vortices and conditional densities. We have also compared the
properties of non-interacting electrons to interacting ones in quantum
dot molecules to separate the effects of the non-circular confinement
potential and interactions.

\begin{acknowledgments}
This work has been supported by the Academy of Finland through its
Centers of Excellence Program (2000-2005). 
\end{acknowledgments}

\end{document}